\documentclass[useAMS,usenatbib]{mn2e}
\usepackage{epsfig}

\title[Eruptive YSOs]{A systematic survey for eruptive young stellar objects using mid-infrared
photometry}

\author[Scholz et al.]{Alexander Scholz$^{1}$\thanks{E-mail: aleks@cp.dias.ie}, 
Dirk Froebrich$^{2}$, Kenneth Wood$^3$\\
$^{1}$ School of Cosmic Physics, Dublin Institute for Advanced Studies, 31 Fitzwilliam Place, 
Dublin 2, Ireland\\
$^{2}$Centre for Astrophysics and Planetary Science, University of Kent, Canterbury, CT2 7NH, United Kingdom\\
$^{3}$School of Physics and Astronomy, University of St. Andrews, The North Haugh, St. Andrews, Fife, KY16 9SS, United Kingdom\\
}

\begin{document}

\date{Accepted. Received.}

\pagerange{\pageref{firstpage}--\pageref{lastpage}} \pubyear{2002}

\maketitle

\label{firstpage}

\begin{abstract}
Accretion in young stellar objects (YSOs) is at least partially episodic, i.e. periods with high 
accretion rates ('bursts') are interspersed by quiescent phases. These bursts manifest 
themselves as eruptive variability.
Here we present a systematic survey for eruptive YSOs aiming to constrain the frequency of accretion 
bursts. We compare mid-infrared photometry from Spitzer and WISE separated by $\sim 5$\,yr for two samples 
of YSOs, in nearby star forming regions and in the Galactic plane, each comprising about 4000 young sources. 
All objects for which the brightness at 3.6 and 4.5$\,\mu m$ is increased by at least 1\,mag between the 
two epochs may be eruptive variables and burst candidates. For these objects, we carry out follow-up 
observations in the near-infrared. We discover two new eruptive variables in the Galactic plane which 
could be FU Ori-type objects, with K-band amplitudes of more than 1.5\,mag. One object known to undergo
an accretion burst, V2492 Cyg, is recovered by our search as well. In addition, the young star ISO-Oph-50, 
previously suspected to be an eruptive object, is found to be better explained by a disk with varying 
circumstellar obscuration. In total, the number of burst events in a sample of 4000 YSOs is 1-4. Assuming 
that all YSOs undergo episodic accretion, this constraint can be used to show that phases of strong accretion 
($>10^{-6}\,M_{\odot}$yr$^{-1}$) occur in intervals of about $10^4$\,yr, most likely between 5000 and 50000\,yr. 
This is consistent with the dynamical timescales for outflows, but not with the separations of emission 
knots in outflows, indicating that episodic accretion could either trigger or stop collimated
large-scale outflows. 
\end{abstract}


\section{Introduction}
\label{s0}

Accretion flows from a circumstellar disk onto a young stellar object (YSO) play a key role in the early
evolution of objects over a wide range of masses, from massive Herbig Ae/Be stars to brown dwarfs.
Observations suggest that the accretion process is non-steady, with episodic bursts with high 
rates of mass accretion interspersed by significantly longer quiescent phases. The evidence for 
episodic accretion rests on three findings: 1) the fact that the luminosities of most protostars are
dominated by internal radiation, not by heating due to accretion \citep[e.g.][]{2009ApJS..181..321E}; 
2) the discontinuities seen in protostellar outflows, which constitute a fossil record of the accretion history 
\citep[e.g.][]{2012arXiv1206.5095I}; 3) the discovery of a small number of eruptive variables which are currently 
experiencing strongly enhanced accretion rates with respect to the typical YSOs, with FU Ori as the prototype 
\citep[e.g.][]{1996ARA&A..34..207H,2010vaoa.conf...19R}. 

While the general idea of episodic accretion is well-established, the driving force of the bursts
is not known yet. In general, these events are explained in the framework of various disk instabilities,
e.g. thermal instabilities \citep[see][and references therein]{1994ApJ...427..987B}, gravitational
instabilities \citep{2005ApJ...633L.137V,2012ApJ...747...52D}, or different types of magnetic
instabilities \citep{2001MNRAS.324..705A,2011ApJ...740L...6M,2009ApJ...701..620Z}. In addition,
various types of trigger events are discussed in this context, e.g. star-star encounters \citep{2010MNRAS.402.1349F}, 
star-disk encounters \citep{2008A&A...492..735P}, tidal effects from a companion star \citep{1992ApJ...401L..31B}, 
or interactions between the disk and a massive planet \citep{2004MNRAS.353..841L,2005MNRAS.361..942C}. 
These various scenarios lead to specific predictions regarding the frequency and properties of bursts. 

Strong accretion bursts may also be a relevant factor in the context of planet formation and could
have an impact on the architecture and frequency of planetary systems. For example, the 
length of the 'lulls' between bursts may limit the efficiency of planet formation via disk
fragmentation \citep{2011ApJ...730...32S}. FU Ori-type bursts, caused by gravitational instabilities,
have also been suggested as events that provide the transient shock heating needed to explain the 
formation of chondrules \citep[e.g.][]{2008ApJ...685.1193B}. 

In this context, it may be useful to see accretion eruptions as a weather-like phenomenon in the disk 
('disk weather'): a process that affects the physics of the disk, but is to some extent random and 
occurs on timescales that are extremely short compared with the disk lifetime. 
Observational studies on large samples are essential to constrain the characteristics of this 
process and to guide the theoretical work. So far, however, most FU Ori-type and other bursts have
been found serendipitously, which does not allow to put rigorous constraints on their frequency.
The advent of wide-area, infrared surveys of large numbers of star forming regions makes systematic 
surveys for accretion bursts feasible. In this paper we present the results from such a survey. 
The goal is to derive an estimate of the frequency of bursts using two epochs of mid-infrared photometry 
provided by the Spitzer and the WISE satellites. We aim to probe the largest sample of YSOs that 
is available for such a comparison, in total about 8000 objects covering a wide range of masses and ages. 

\section{The approach}
\label{s1}

\subsection{The data}

We aim to constrain the frequency of accretion bursts by comparing two epochs of mid-infrared
photometry from Spitzer and WISE \citep{2010AJ....140.1868W}. Two of the channels used by these 
satellites can be compared with each other: IRAC1 and WISE1 with central wavelengths at 3.4-3.6$\,\mu m$ 
as well as IRAC2 and WISE2 at 4.5-4.6$\,\mu m$. The differences in these two bands between the two 
telescopes are minor and can be neglected here as we are only interested in variability with large 
amplitudes. The epoch difference between the Spitzer and WISE observations depends on the area of the 
sky; the samples of YSOs used in this paper have been observed by Spitzer between 2003 and 2006, 
whereas most of the WISE data has been taken in 2010. Thus, the epoch differences in our samples 
are 4-7\,yr, with a typical value of $\sim 5$\,yr.

\subsection{A simplified model for episodic accretion}
\label{model}

With two epochs we can only test for a specific type of episodic accretion. We will search for objects 
undergoing a burst event with a rise time $t_1$ and a decline time $t_2$, where $t_1$ and $t_2$ are assumed 
to be shorter and longer than our typical epoch difference of 5\,yr, respectively. We 
also assume that any additional variability in YSOs is small compared with the events caused by the accretion 
bursts. These conditions are fulfilled for most, but not all, of the known FU Ori type objects. 
One exception is V1515 Cyg, one of the best-studies FU Ori objects, which exhibits a long rise time 
of about 20\,yr \citep{2005MNRAS.361..942C}. In general, the known FU Oris show considerable diversity in their 
lightcurves which is not represented in this simple model. The quantity we are aiming to constrain is the 
{\it typical interval between consecutive bursts}. According to previous estimates, this interval is in 
the order of several thousands of years and thus much longer than the typical duration of a burst 
\citep{1977ApJ...217..693H,1996ARA&A..34..207H}.

When comparing two epochs of photometry, the burst interval can be crudely estimated as $I = \Delta t \times N / n_B$. 
Here $\Delta t$ is the epoch difference between the two observations, $N$ the sample size, $n_B$ the
number of detected bursts in that sample, and $I$ the desired quantity.
This simple relation serves as a useful starting point for the analysis; for a more accurate statistical
evaluation we will use Monte-Carlo simulations (Sect. \ref{s4}). It is clear that maximum information
can be gained by maximising the sample size and the epoch difference. For our study the epoch difference
is fix, i.e. the key is to make the sample as large as possible. For example, with $\Delta t = 5$\,yr, 
we need in the order of 1000 objects to have a substantial chance of detecting at least one event, if the
interval between bursts is 5000\,yr. Based on the expected intervals, we thus need to cover several 
thousands of young stars to be able to provide useful limits.

In the literature the quantity that is often used to describe episodic accretion is the 'duty cycle', i.e.
the fraction of time a YSO spends in the FU Ori state. Measuring the duty cycles requires knowledge of the duration 
of accretion bursts. Since the slow decline is much more difficult to constrain from direct observations
than the fast rise of an accretion burst, we focus here on the burst interval rather than the duty cycle.

We note that with our approach we do not make an attempt to distinguish between the various types of 
accretion bursts presented in the literature, with FU Oris as the most extreme examples and EXors as
smaller events \citep[see][]{2010vaoa.conf...19R}. We are simply interested in any type of eruptive event 
in a YSO, which could be due to an increase in mass accretion rate.

\subsection{Flux increase during accretion bursts}

Objects undergoing an accretion burst manifest themselves as eruptive variables with strongly increased luminosities 
at all optical and infrared wavelengths. Assuming that all the gravitational energy from infalling material is converted 
to radiation, the additional luminosity from an accretion rate of $10^{-6}\,M_{\odot}$yr$^{-1}$, exceeds the solar 
luminosity by more than one order of magnitude (factor 15, assuming a star with $M = 1\,M_{\odot}$ and $R = 2\,R_{\odot}$). 
To evaluate how this additional energy is distributed across the
spectrum, we used the Monte Carlo radiative transfer models discussed in detail in \citet{2006ApJ...645.1498S}
(see also \citet{2006ApJS..167..256R} for more information). In short, the code is based on the following
assumptions: 1) NextGen stellar atmospheres are used for the photospheric spectrum; 2) the grain size distribution
in the disk follows a power law with an exponential decay for particles with sizes above 50$\,\mu m$ and a formal
cutoff at 1\,mm; 3) dust in regions close to the star is destroyed if the temperature is above the
dust sublimation threshold; 4) the scaleheight of the disk increases with radius following $h(r) = h_0 (r/R_{\star})^\beta$;
5) the accretion luminosity is split between disk and star, where the stellar part is distributed evenly over
the stellar surface (i.e. no hot spots).

\begin{figure*}
\includegraphics[width=18.0cm,angle=0]{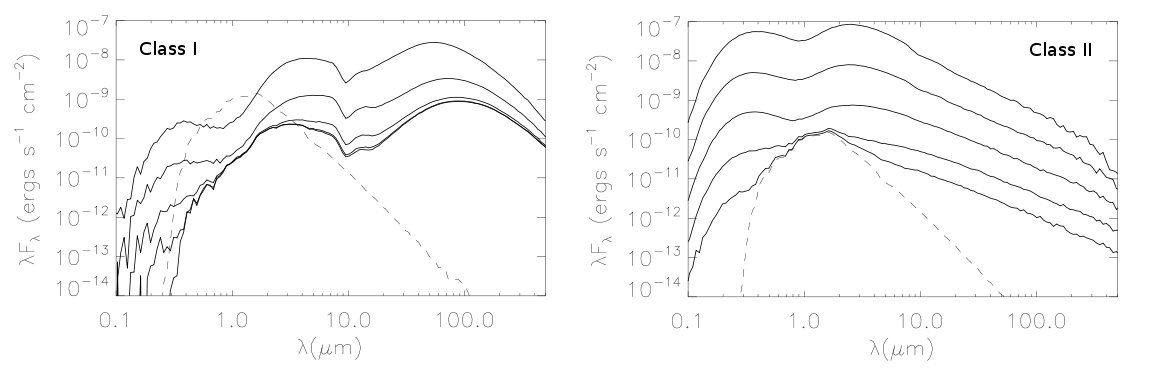}
\caption{Spectral energy distributions from radiative transfer modeling for prototypical YSOs with varying 
accretion rates. Left panel: 'Class I prototype' with disk and massive envelope. Right panel: 'Class II prototype'
with disk only. For each prototype, 5 SEDs are shown for accretion rates of $10^{-5}$, $10^{-6}$, $10^{-7}$, $10^{-8}$
and $0.0\,M_{\odot}$yr$^{-1}$ (from top to bottom). The dashed line is the photospheric SED.
\label{f0}}
\end{figure*}

For the purposes of this paper, we do not aim to explore in detail the parameter space; instead we want to
find a typical value for the flux increase in a given wavelength domain as a function of accretion rate. We also
neglect the fact that a strong increase in the mass accretion rate will affect the structure of the disk.
In Fig. \ref{f0} we show model SEDs for a prototypical Class I source (stellar mass 0.5$\,M_{\odot}$, disk mass 
0.1$\,M_{\odot}$, envelope mass 2.0$\,M_{\odot}$) and for a prototypical Class II source (stellar mass 0.5$\,M_{\odot}$, 
disk mass 0.01$\,M_{\odot}$, no envelope), for a range of accretion rates from 0.0 to $10^{-5}\,M_{\odot}$yr$^{-1}$. 
These figures illustrate that the flux increase at 2-5$\,\mu m$ compared with the zero accretion case is around one 
order of magnitude for accretion rates of $10^{-6}\,M_{\odot}$yr$^{-1}$ or larger. 

Fig. \ref{f0} warrants two additional comments: 1) At all accretion rates, the amplitude is substantially larger 
for the Class II prototype. This is caused by the presence of an envelope in the Class I system, combined with its
high photospheric luminosity, which is due to the inflated radius of the central source. The additional infrared flux 
from the envelope, heated by a brighter central source, 'drowns' the contribution from the accretion, i.e. the relative 
flux increase due to accretion is smaller than in the Class II stage.
2) The model with the lowest fluxes corresponds to a (theoretical)
accretion rate of zero. In practise, this model is indistinguishable from models with $10^{-9}\,M_{\odot}$yr$^{-1}$
or lower, values which are frequently measured for T Tauri stars \citep{2006A&A...452..245N}. This illustrates that for
most T Tauri stars accretion does not contribute significantly to the mid-infrared flux.

This exercise suggests that Class I and II sources whose accretion rate increases to $10^{-6}\,M_{\odot}$yr$^{-1}$
or more are expected to increase in brightness by at least 2.5\,mag at near/mid-infrared wavelengths. In contrast,
the {\it typical}, short-term, near/mid-infrared variations in large samples of YSOs, due to rotation, hot spots, and inner disk
inhomogenities, are in the range of 0.1-0.6\,mag \citep{2011ApJ...733...50M,2012ApJ...748...71F}. In our
Spitzer-WISE comparison we will therefore adopt a cutoff of 1.0\,mag to select burst candidates. On one hand, this
should avoid most of the other types of variability in these sources; on the other hand it should also select
eruptions where the two epochs of photometry do not catch the maximum and minimum.

\section{Identification of burst candidates}
\label{s2}

In the following section, we will discuss the selection of possible eruptive variables and thus burst candidates
from archival Spitzer and WISE photometry, as well as the follow-up observations and their results. 

\subsection{The C2D catalogue}
\label{c2d}

The 'Cores to Disks' (C2D) Spitzer legacy program has provided a catalogue of YSO candidates for nearby
molecular clouds and small cores, identified using near- and mid-infrared colour criteria \citep{2009ApJS..181..321E}. 
In total, the catalogue comprises 1478 sources from the subsamples CLOUDS, OFF-CLOUD, CORES, and 
STARS. We obtained this list from IPAC and searched for matches in the WISE all-sky catalogue \citep{2010AJ....140.1868W}. 
For 1323 objects a match was found within 2", for the overwhelming majority of them the distance between Spitzer and 
WISE coordinates is well below 1". 1301 of these objects have a robust detection in the Spitzer and WISE channels at 
3.6 and 4.5$\,\mu m$ (signal-to-noise ratio $>5$ for WISE, error $<20$\% for Spitzer). 1296 of these are also 
robustly detected in the Spitzer channels at 5.8 and 8.0$\,\mu m$. 

\begin{figure*}
\includegraphics[width=6.6cm,angle=-90]{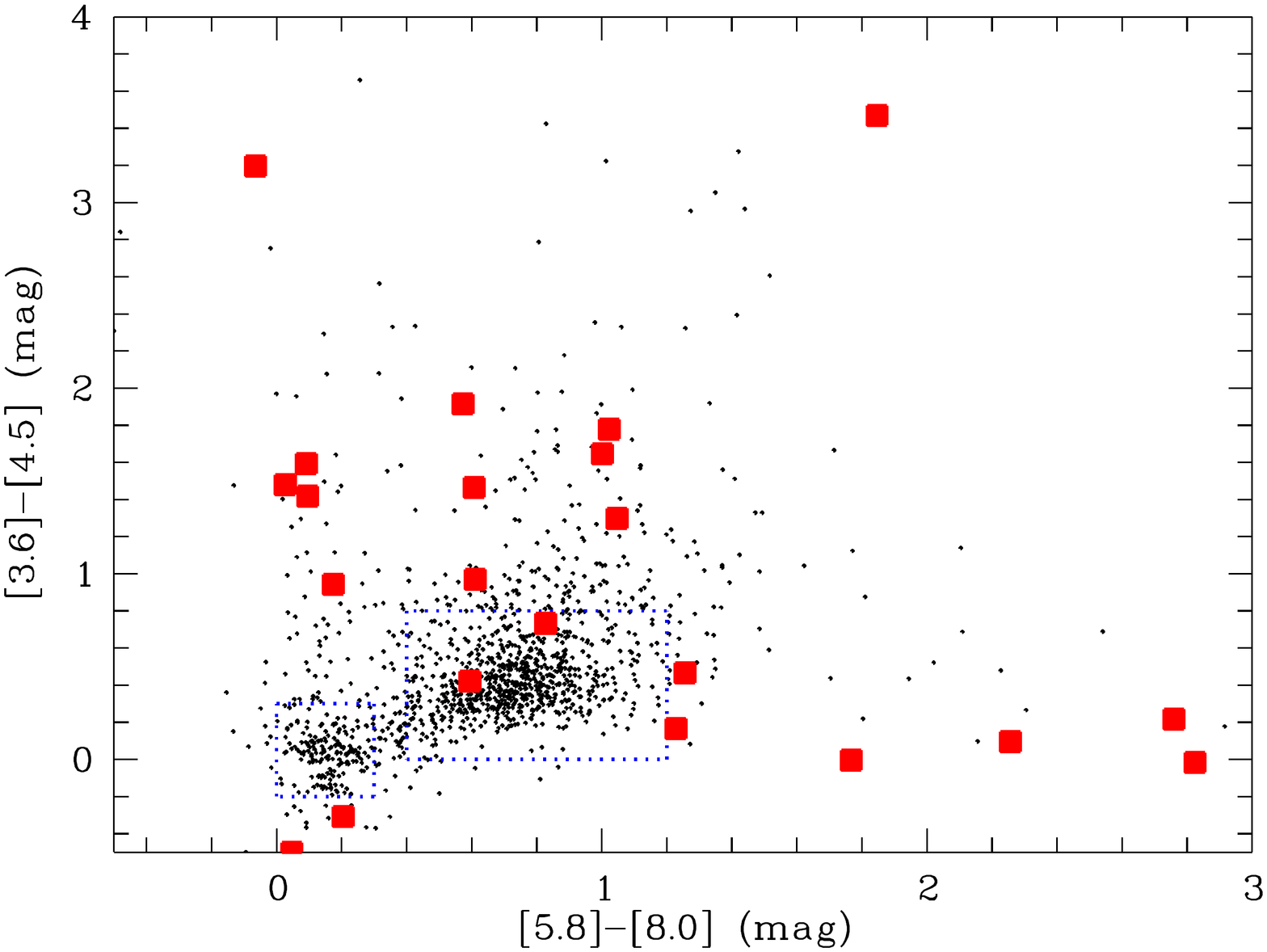}
\includegraphics[width=6.6cm,angle=-90]{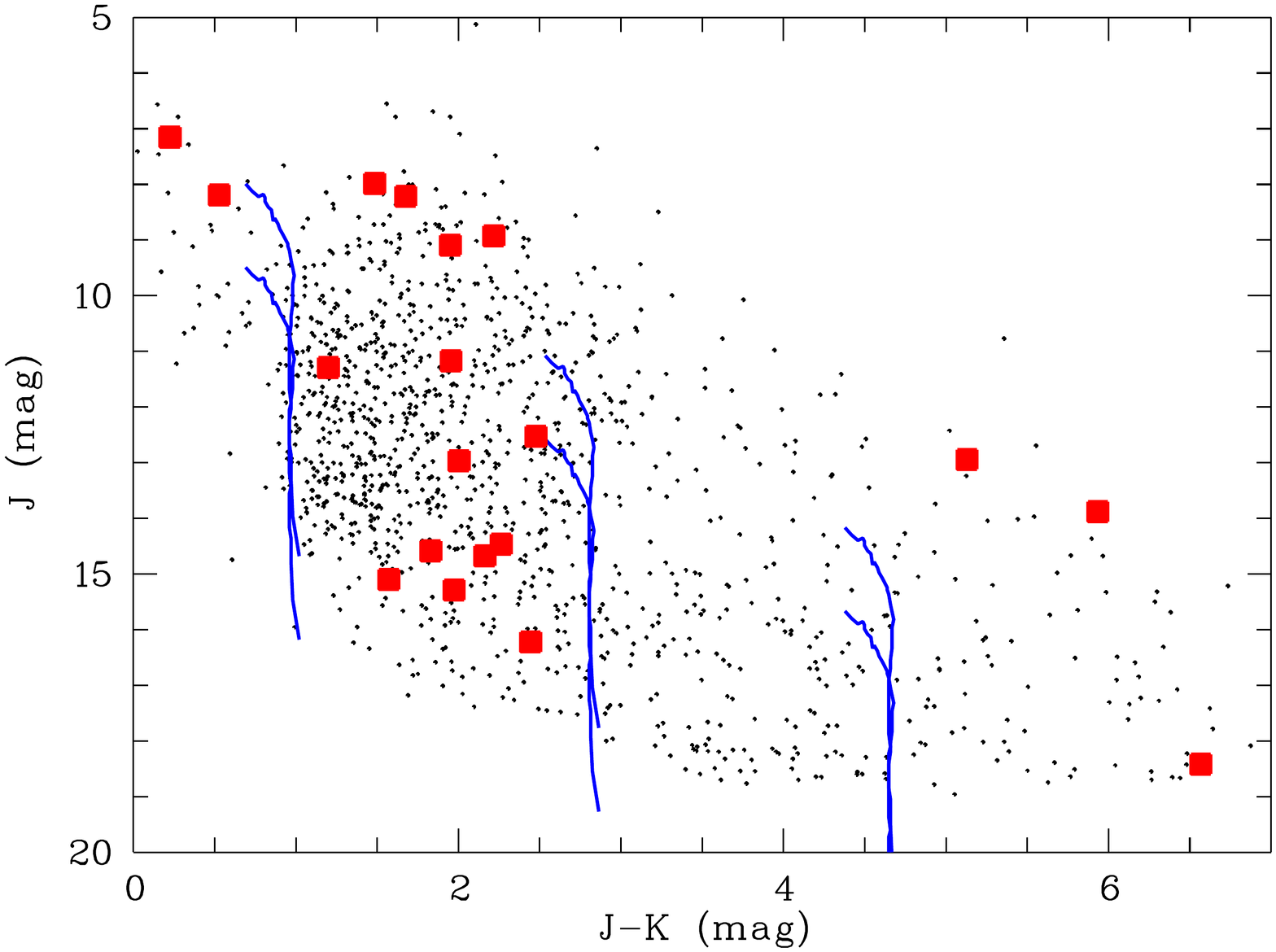}
\caption{Colours of objects in the C2D sample. Burst candidates are marked with large red squares.
Left panel: IRAC colour-colour plot for all objects with robust detections in all 4 IRAC 
channels (1296 out of the total sample of 1301). The Class III and Class II locus are 
shown as dotted blue boxes around the origin and right of the origin. Right panel: Near-infrared 
colour-magnitude diagram for the subsample with 2MASS photometry (1228 objects). BCAH 
isochrones for an age of 1\,Myr, distances of 150 and 300\,pc, and extinctions of 
$A_V = 0$, 10, 20\,mag are overplotted. 
\label{c2dfig}}
\end{figure*}

In Fig. \ref{c2dfig} (left panel) we show the IRAC colour-colour plot for this sample. Two cumulations, 
around the origin and right of the origin of the diagram, are clearly seen and can be identified as 
the locus of the Class III (no disk) and II sources. The sample contains 115 objects with typical Class III 
colours (around the origin) and 324 with typical Class II colours (right of the origin). 249 objects 
are above the Class II box, which makes them good candidates for embedded Class I sources. The 
remaining sources are scattered around these areas. According to \citet{2009ApJS..181..321E}, about one 
third of the CLOUDS subsample are in the early embedded stage (Class I or Flat).

In addition, Fig. \ref{c2dfig} (right panel) shows the (J,J-K) colour magnitude diagram for the 1228 
sources with 2MASS near-infrared photometry in the sample, to assess the properties of the central 
sources. Overplotted are the BCAH 1\,Myr isochrones \citep{1998A&A...337..403B} which range from 0.02 to 
1.4$\,M_{\odot}$, for distances of 150\,pc and 300\,pc, bracketing the regions covered by C2D, and 
for $A_V=0$, 10, and 20\,mag. The comparison with the models illustrates that 
the sources cover the low-mass regime down to the substellar limit, including brown dwarfs at low 
extinctions, but only relatively few objects with $M>1.4\,M_{\odot}$. About two thirds to three 
quarters of the sample have extinctions below $A_V=10$\,mag.

\begin{figure*}
\includegraphics[width=6.6cm,angle=-90]{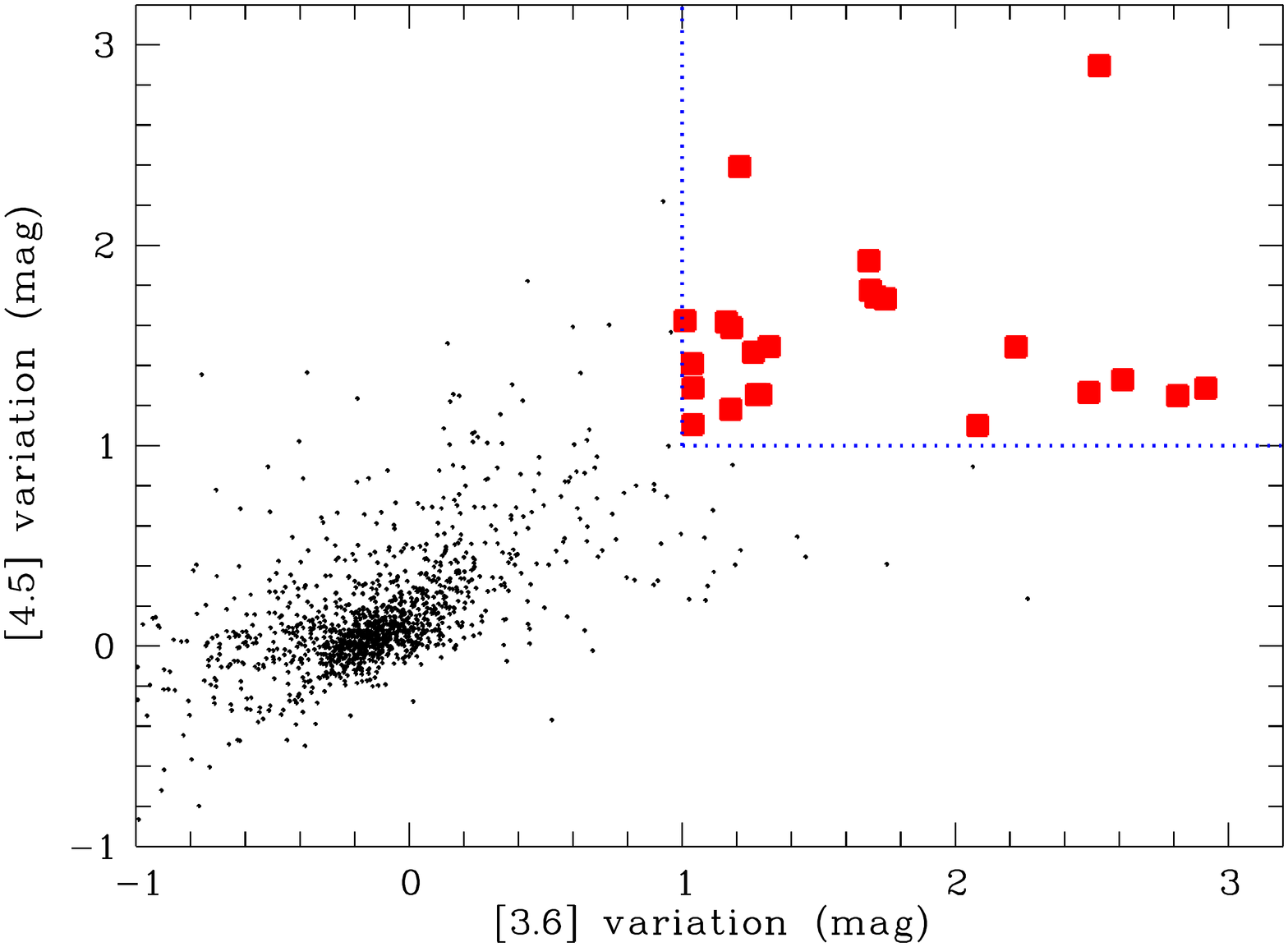}
\includegraphics[width=6.6cm,angle=-90]{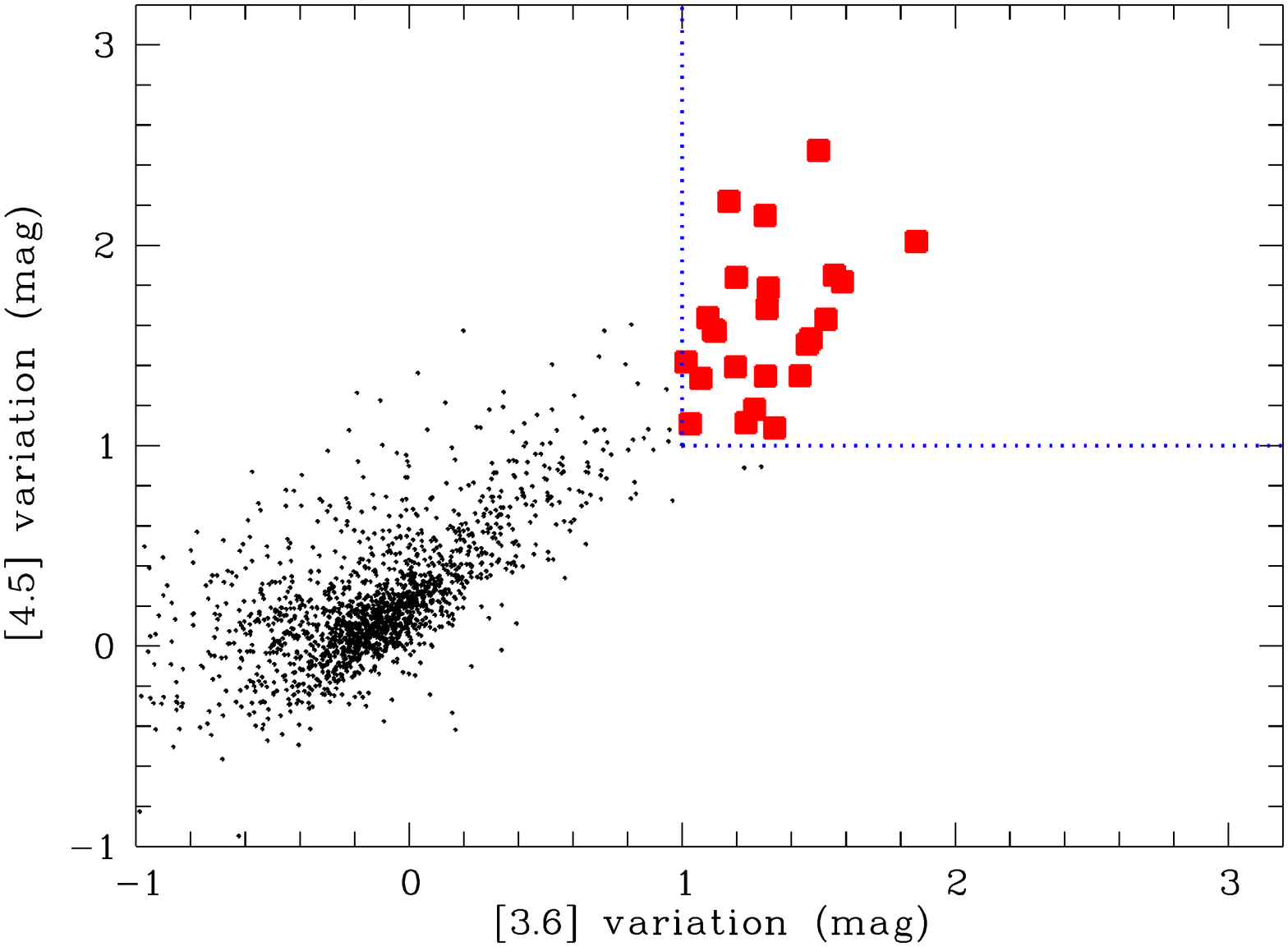}
\caption{Variability in C2D sample (left) and Cluster (right) sample. The variations are calculated as 
difference between C2D magnitudes and WISE magnitudes, i.e. positive values indicate a brightening. 
Objects in the upper right corner (large symbols) show a brightness increase by more than 1\,mag in the 
two mid-infrared bands.
\label{c2dclustvar}}
\end{figure*}

In Fig. \ref{c2dclustvar} (left panel) we illustrate the selection of variables from this sample. The 
differences in the magnitudes at 3.6 and 4.5\,$\mu m$ between C2D and WISE show a clear cumulation around 
(0,0), as expected, because these bandpasses of IRAC and WISE are comparable. The objects of interest to 
us are located in the upper right corner. 23 sources are more than 1\,mag brighter in WISE compared 
with C2D in the two bands, providing evidence for a substantial increase in the brightness. These 
objects are also overplotted in Fig. \ref{c2dfig}, as far as they have the required photometry (22 
in the left panel, 20 in the right panel). They do not show an obvious bias in the (J,J-K) diagram, 
but most of them are above the Class II locus in the IRAC colour-colour plot, indicating that 
they may be embedded Class I sources. 

All 23 highly variable sources were checked individually in the available images from WISE, Spitzer,
and 2MASS. 5 of them are galaxies in 2MASS images and can be ruled out. For the remaining we 
obtained the C2D and WISE images at 3.6 and 4.5\,$\mu m$ and compared them. In at least 4 cases the 
flux increase in the WISE catalogue can be attributed to close neighbours that were not resolved 
with WISE, due to its significantly broader PSF (6" vs. 2", see \citet{2010AJ....140.1868W}). For 8 
others, the IRAC photometry is affected by saturation. 4 more are extended objects in the IRAC images 
and could be part of a protostellar outflow. For the remaining 2, no obvious reason for the flux 
difference in the C2D and WISE catalogues can be identified, but the images clearly show that the 
object did not become significantly brighter. Thus, none of the candidates from the C2D sample 
classifies as a burst candidate.

\subsection{The Cluster catalogue}
\label{sampleb}

The second sample is derived from the catalogue of YSOs in clusters within 1\,kpc published
by \citet{2009ApJS..184...18G}. The list of 2548 objects has been selected based on Spitzer/IRAC
and MIPS data using mid-infrared colour cuts. It covers 36 nearby clusters, star forming clouds,
and young groups, including some overlap with the regions covered in the C2D sample. We obtained
the catalogue from Vizier and cross-matched with the WISE database. 1796 objects have a WISE
match within 2", 1672 of them within 1". 1745 have a robust detection (criteria as above) in
the 3.6 and 4.5$\,\mu m$ channels of Spitzer and WISE, 1587 of them with additional data in the 
J- and K-band from 2MASS, 1642 of them with data in the two IRAC channels at 5.8 and 8.0$\,\mu m$.
Note that 380 objects from the Cluster sample are also contained in the C2D sample. 

\begin{figure*}
\includegraphics[width=6.6cm,angle=-90]{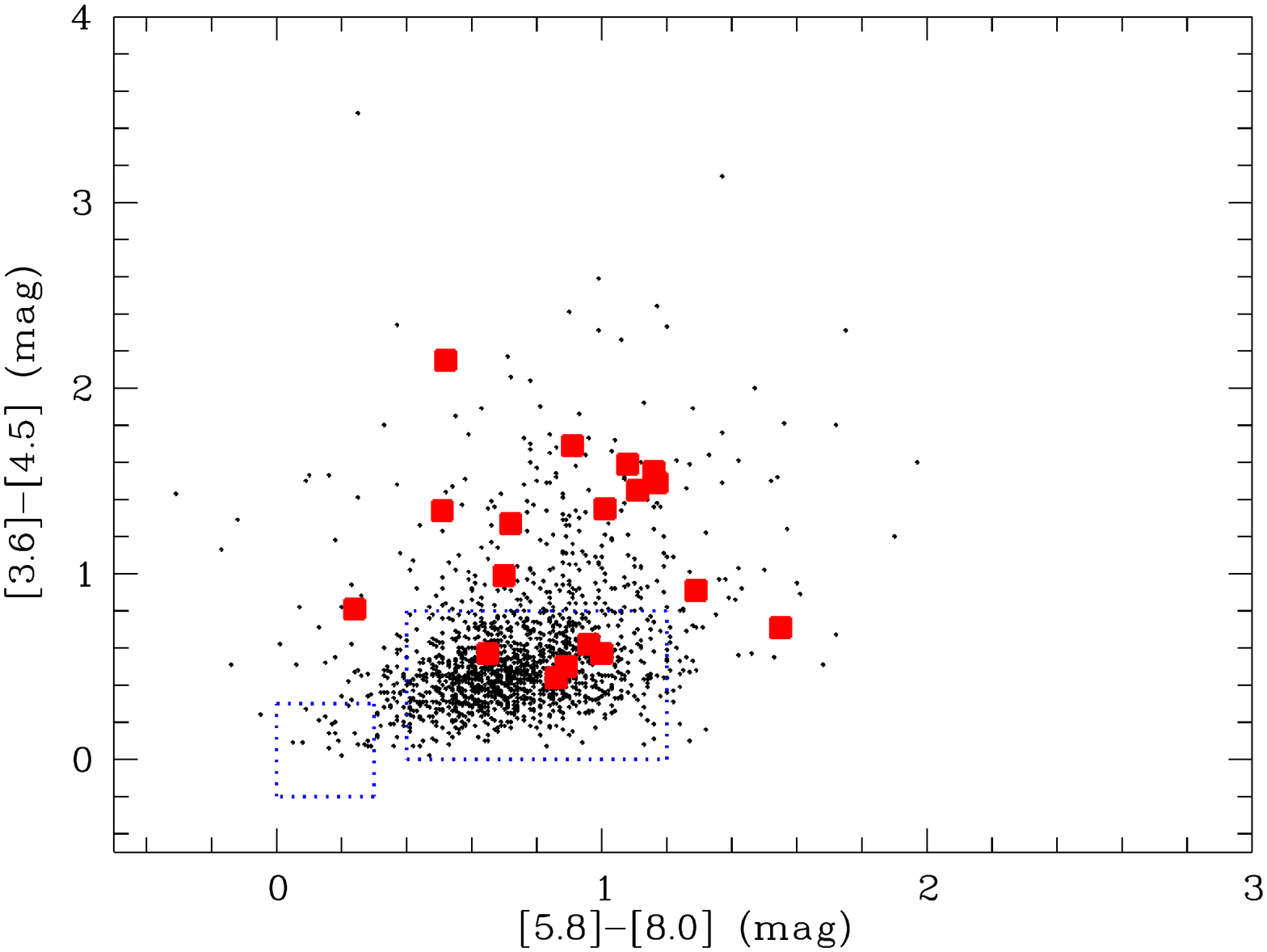}
\includegraphics[width=6.6cm,angle=-90]{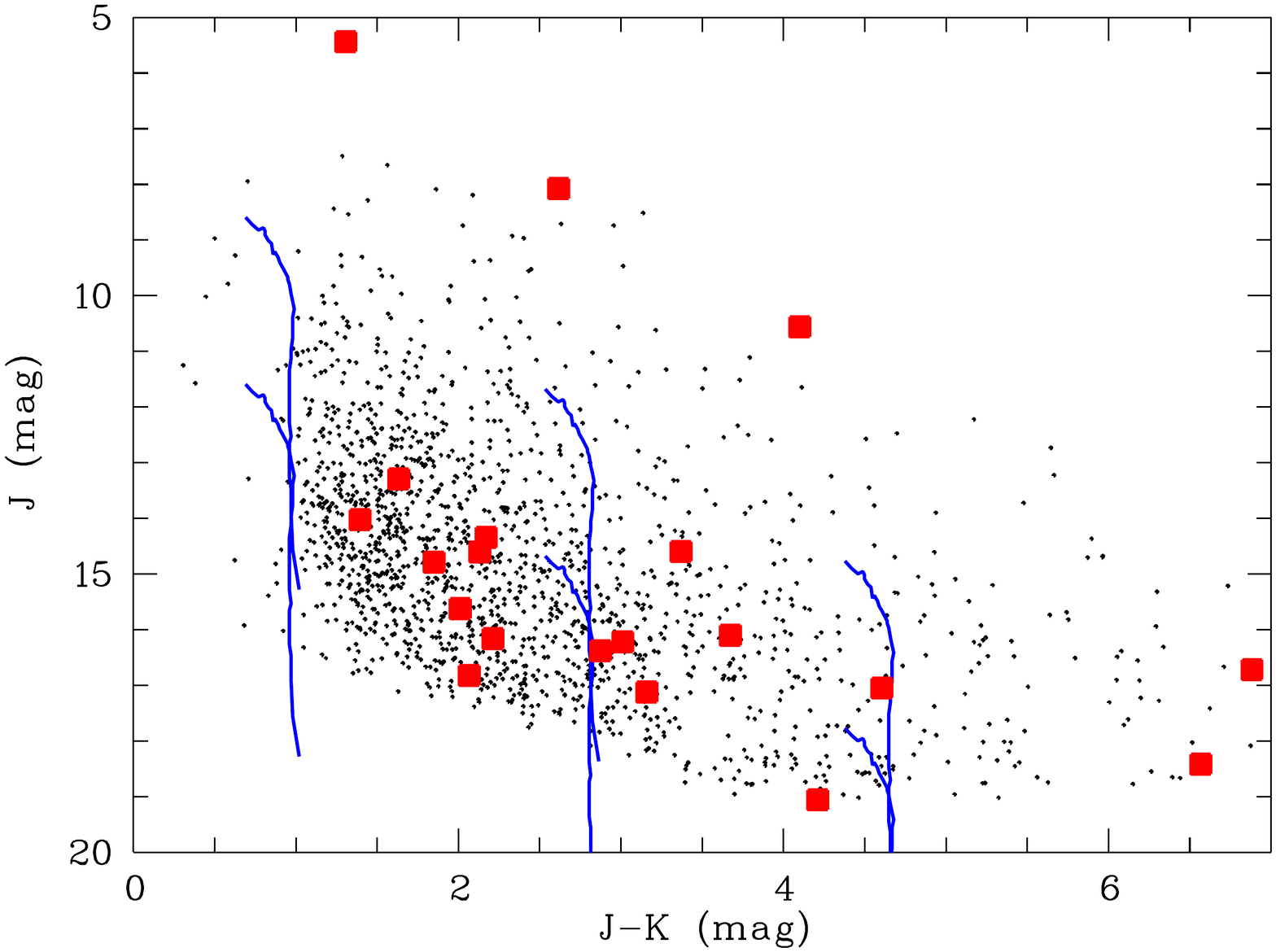}
\caption{Colours of objects in the Cluster sample. Burst candidates are marked with large symbols.
Left panel: IRAC colour-colour plot for all objects with robust detections in all 4 IRAC 
channels (1642 out of the total sample of 1745). The Class III and Class II locus are 
shown as dotted blue boxes around the origin and right of the origin. Right panel: Near-infrared 
colour-magnitude diagram for the subsample with 2MASS photometry (1587 objects). BCAH 
isochrones for an age of 1\,Myr, distances of 200 and 800\,pc, and extintions of 
$A_V = 0$, 10, 20\,mag are overplotted. 
\label{clust}}
\end{figure*}

As for the C2D sample, we show the IRAC colour-colour plot and the (J, J-K) near-infrared 
colour-magnitude diagram for this sample in Fig. \ref{clust}. In contrast to the C2D sample,
the Cluster objects do not contain a significant fraction of Class III sources, the majority
is classified as Class II. Based on our diagram, we estimate that at least 1226 out of 1642 are
Class II (75\%), the classification provided by \citet{2009ApJS..184...18G} yields an even higher
fraction of 86\%. About 15-20\% objects in this sample are Class I.

In the near-infrared plot we show the 1\,Myr BCAH isochrones for distances of 200 and 800\,Myr, bracketing
most of the objects in the sample, for three different extinctions. The plot demonstrates that the sample
is dominated by low-mass stars at extinctions of $A_V<20$\,mag. The sample includes substellar objects,
but only for the regions with distances $<500$\,pc and low extinctions. In general, the characteristics
of this sample make it comparable to the C2D sample.

Fig. \ref{c2dclustvar} (right panel) shows the variability in the Cluster sample. 24 objects fulfill our 
variability criterion and have an increased brightness by $>1.0$\,mag in the WISE catalogue in the two bands. 
One of these objects has already been identified in the C2D sample. As before, these burst candidates
were checked in the available images. Seven of them have close, usually brighter neighbours, which may have
caused an apparent brightness increase. For 7 others the brightness in WISE is probably affected by the 
surrounding nebulosity. Four more are saturated in the Spitzer/IRAC images. We are left with 5 candidates 
which appear to be brighter in the WISE images and remain burst candidates. One of them, ISO-Oph-50 in the star 
forming region $\rho$-Ophiuchus, has been suspected to be an outbursting young star by \citet{2008A&A...485..155A}, 
due to a brightening by more than 1\,mag over about a year, although it could also be a different type
of variable \citep{2012A&A...539A.151A} -- see discussion in Sect. \ref{comm}.

\subsection{Complementary samples similar to C2D and Cluster}
\label{compl}

We carried out the same test as above in three smaller samples of YSOs, gathered from the literature.

According to \citet{2009ApJS..184...18G}, the Cluster sample covers all clusters within 1\,kpc 
from the \citet{2003ARA&A..41...57L} census, with the exception of NGC2264 and the Orion Nebula Cluster. 
For NGC2264 there is a comprehensive catalogue of the Spitzer photometry available \citep{2009AJ....138.1116S}, 
which allows us to include it in this study. Out of the 490 cluster members identified by 
H$\alpha$ photometry by \citet{2005AJ....129..829D}, 485 have a Spitzer counterpart within 
3". Out of these, 355 have a WISE counterpart within 3" with robust photometry (defined as 
in the other samples). From this list, 5 objects have increased their brightness in the two 
mid-infrared bands by at least 1\,mag. Two of them have little H$\alpha$ emission ($<5$\,\AA) and 
very low IRAC colours (I1-I2\,$<0.1$), which rules out that they harbour a disk. Two sit very close 
to bright stars (or multiple stars) which contaminate their WISE fluxes. One has a nearby equally 
bright neighbour which is not resolved in WISE. To sum up, none of the likely members of NGC2264 is
a burst candidate.

For the nearby star forming region Taurus, \citet{2010ApJS..186..259R} published a census of previously
confirmed members and new candidate members based on Spitzer photometry. Combining their list of known
and new objects and exluding a few without IRAC photometry yields 328 objects from which 236 have previously
been known or have been classified by \citet{2010ApJS..186..259R} as 'most believable'. From this sample
of 328, 320 have robust photometry in the first two WISE bands. Only one of them is more than 1\,mag brighter
in the WISE photometry compared with the Spitzer magnitudes; this object, however, exhibits a 'halo'
and is probably a galaxy.

Another new sample of YSOs from Spitzer data has been published for the various clusters in the
North American and Pelican Nebulae \citep{2011ApJS..193...25R}. Their total sample comprises 1286 IRAC and
MIPS-selected candidate YSOs, about half of them Class II. 1099 have reliable fluxes in the first two IRAC 
and WISE channels. From these, 935 objects have membership are most likely YSOs with flag 'A' or 'B' 
\citep{2011ApJS..193...25R}. This 'A+B' sample may still be affected by significant contamination by AGB
stars, estimated to be between 5 and 25\% by \citep{2011ApJS..193...25R}. Conservatively subtracting
about 20\% reduces the total sample size to about 700.

Four objects fulfill our variability criterium (flux increase by more than 
1\,mag). One of them appears to be extended in the Spitzer images (and has membership flag 'C'), for 
another one the WISE photometry is contaminated by several neighbours. The remaining two are isolated 
and clearly brighter in the WISE images and remain candidates. One of them is the recently identified 
outbursting star V2492 Cyg \citep{2011AJ....141...40C} and has magnitude differences close to 3\,mag 
at 3.6 and 4.5$\,\mu m$. This object became brighter between December 2009 and June 2010 
\citep{2011A&A...527A.133K} and was observed by WISE between June and September 2010, i.e. just 
after the burst. 

Note that this region harbours two more known outbursting stars. The recently identified FU Ori
candidate V2493 Cyg \citep{2011ApJ...730...80M,2011A&A...527A.133K} increased its brightness between 
May and August 2010 and was observed at the end of May with WISE. We find a flux increase by 0.5 and 
0.8\,mag in the mid-infrared channels at 3.6 and 4.5$\,\mu m$, i.e. the Spitzer-WISE comparison may 
have captured the onset of the burst.

The well-known FU Ori star V1057 Cyg with an outburst in 1969 \citep{1971A&A....12..312W} is not 
contained in the \citet{2011ApJS..193...25R} catalogue, presumably due to saturation: The star
is listed in the WISE catalogue with 4.9\,mag at 3.6$\,\mu m$ and -0.3\,mag at 22$\,\mu m$, which
is brighter than the upper limits in the colour-magnitude plots shown by \citet{2011ApJS..193...25R}.

All three additional samples discussed here show similar characteristics to the C2D and Cluster sample 
(similar mass range, similar extinction range, mostly Class II sources). Therefore it is legitimate to 
add them to the C2D and Cluster samples. In total, the sum of C2D, Cluster, NGC2264, Taurus, and North 
American/Pelican Nebulae, minus the objects which appear twice, comprises about 4000 objects, hereafter 
called sample A. This sample yields 7 candidate bursts, out of which 2 have been independently discovered
elsewhere.

\subsection{The Robitaille catalogue}
\label{samplec}

Furthermore, we use the list of intrinsically red sources from
\citet{2008AJ....136.2413R}. This sample contains 18949 objects selected from
the Glimpse\,I and II survey data \citep{2003PASP..115..953B,2009PASP..121..213C}.
\citet{2008AJ....136.2413R} estimate that 50\,\%\,--\,70\,\% of the objects are
YSOs and 30\,\%\,--\,50\,\% are AGB stars. The YSOs in the Robitaille list are
expected to be more distant than 1\,kpc and thus on average more massive than
the sources covered in sample A. The Robitaille catalogue is in the 
following called sample B.

From the full sample we select only objects which have a detection at 3.6 and
4.5\,$\mu$m in Glimpse and WISE, whereby the positions in the two surveys do not
differ by more than one arcsecond. This leaves 12961 targets. To make the sample
as 'clean' as possible, we only consider sources which are brighter than the
completeness limit in this sample (11.5\,mag at 3.6\,$\mu$m, 11.0\,mag at
4.5\,$\mu$m) in both surveys. Here the completeness limit was determined as the
peak in the 3.6 and 4.5\,$\mu$m magnitude distribution. 

One potential issue of this sample is the high stellar density in the Galactic
plane. Since the WISE survey has a larger point spread function than Spitzer,
the presence of bright neighbour stars can cause an apparent increase in the
brightness, when the two surveys are compared. To account for that, we exclude
all objects that have a nearby Glimpse source (within 6" of the Robitaille
object) that is bright enough to cause an increase of more than 0.1\,mag in
either the 3.6 or 4.5\,$\mu$m filter. This final sample contains 7101 objects,
which are, as mentioned above, a mix of YSOs and AGB stars.

\begin{figure*}
\includegraphics[width=6.6cm,angle=-90]{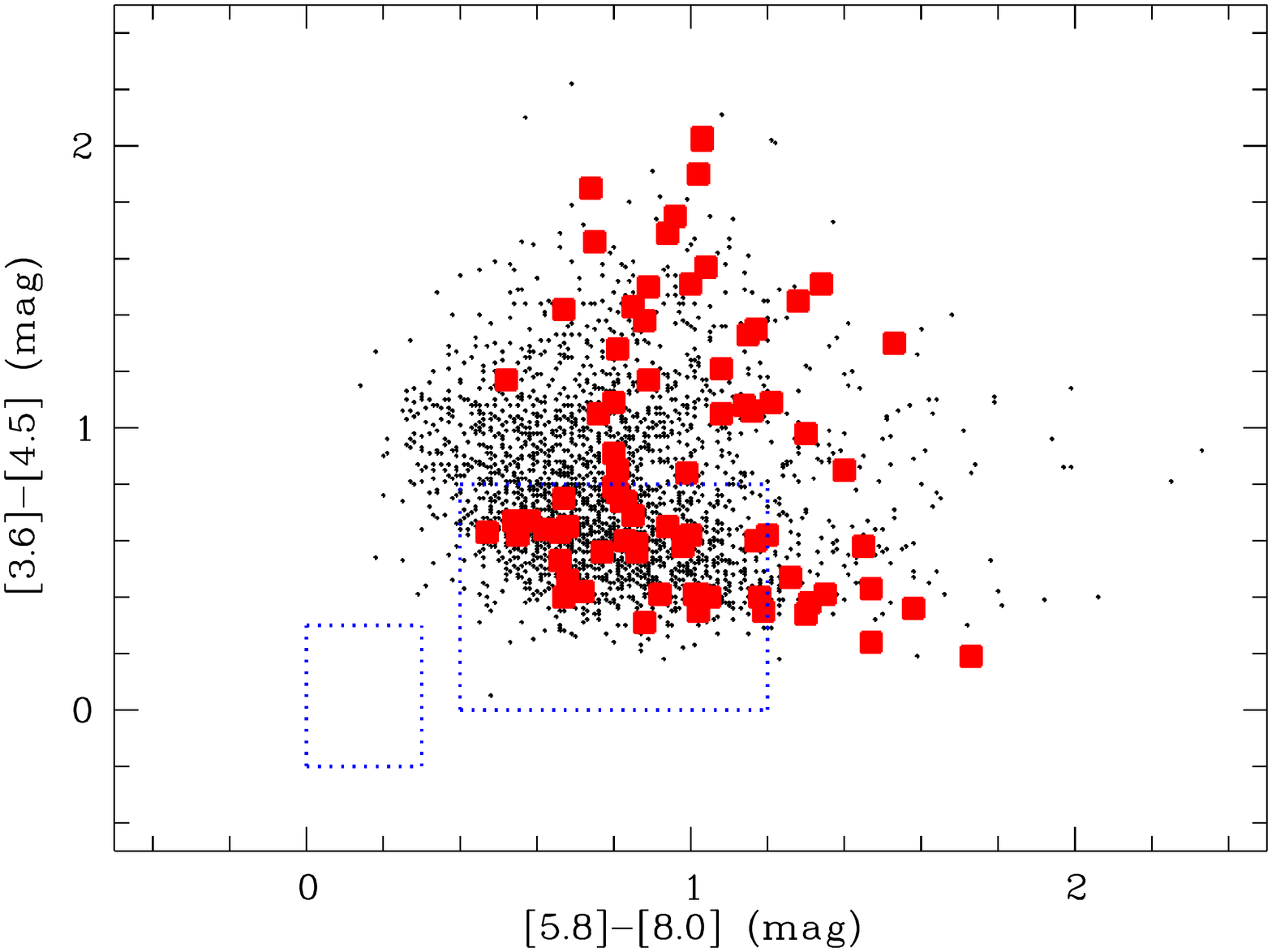}
\includegraphics[width=6.6cm,angle=-90]{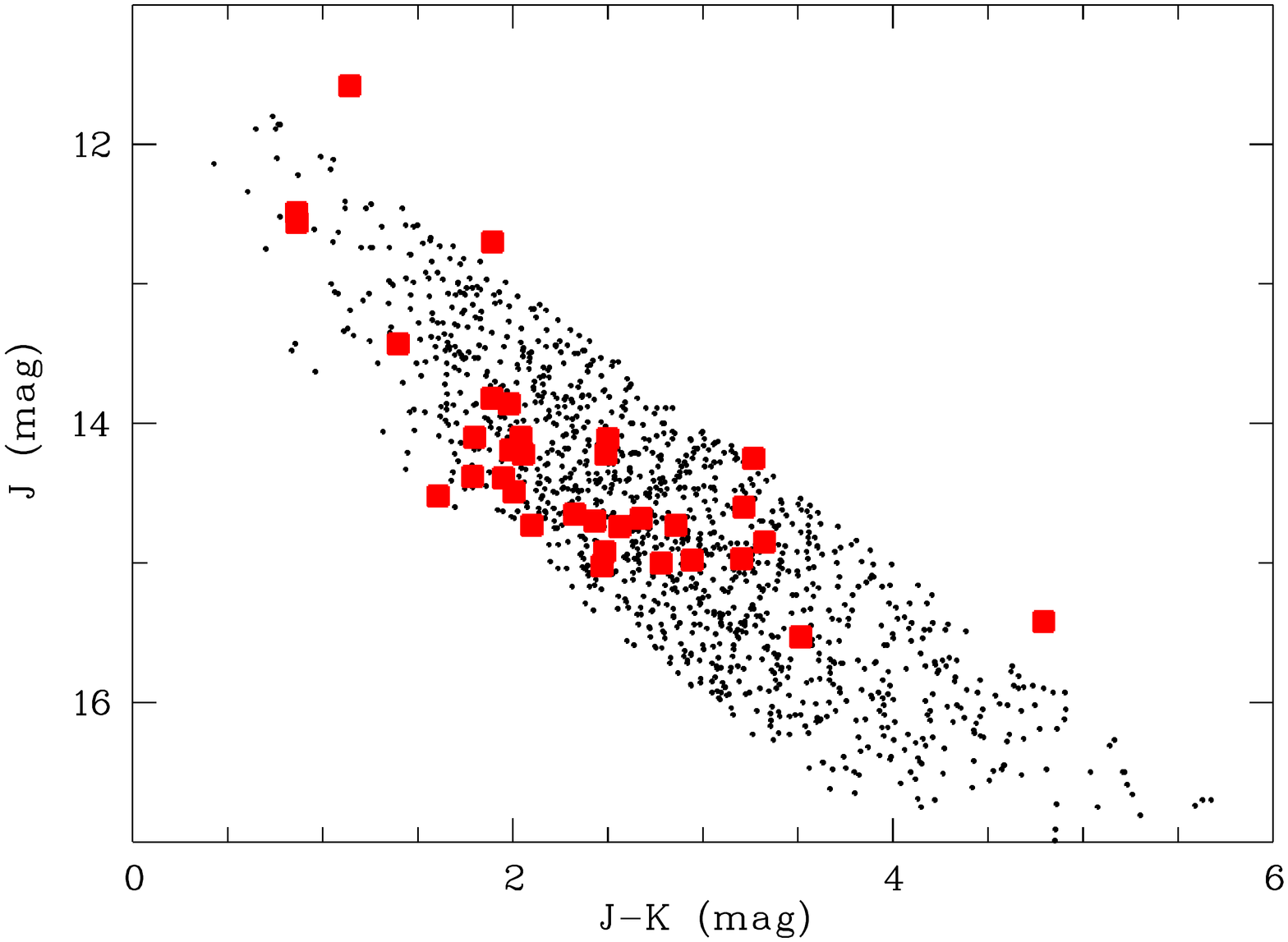}
\caption{\label{rob} Colours of objects in the Robitaille sample (only subsamples YSO1 and YSO2, see Appendix 
\ref{a1}). Burst candidates are marked with large symbols. Left panel: IRAC colour-colour plot, the Class 
III and Class II locus are shown as blue dotted boxes around the origin and right of the origin. Right panel: 
Near-infrared colour-magnitude diagram for the subsample with 2MASS photometry.}
\end{figure*}

In Fig. \ref{rob} we show the usual IRAC colour-colour plot and (J,J-K) colour-magnitude
diagram for a subsample of the Robitaille catalogue that is most likely to be dominated
by YSOs (see Appendix \ref{a1} on the selection of this subsample). As in the other samples, 
most of the sources can be considered Class II based on their mid-infrared colours. As 
expected, the typical $J-K$ colours are larger than in sample A, indicating higher
extinction. 

\begin{figure}
\includegraphics[width=6.6cm,angle=-90]{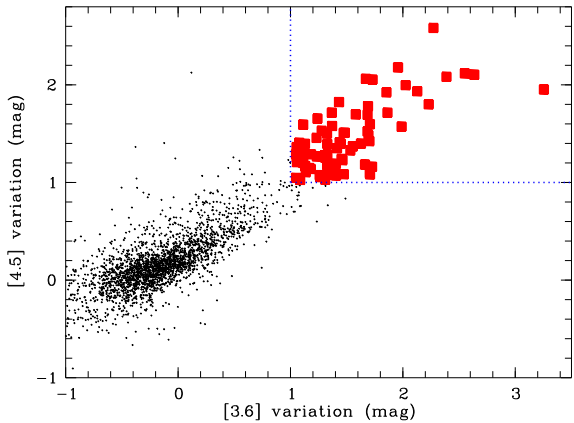}
\caption{\label{robvar} Variability in Robitaille sample (only subsamples YSO1 and YSO2, see Appendix \ref{a1}). 
The variations are calculated as difference between C2D magnitudes and WISE magnitudes, i.e. positive values 
indicate a brightening. Objects in the upper right corner (large symbols) show a brightness increase by more 
than 1\,mag in the two mid-infrared bands.}
\end{figure}

Out of these 7101 sources, there are 77 objects which increase their brightness
by more than one magnitude at 3.6 and 4.5\,$\mu$m (see Fig. \ref{robvar}) and
are possible candidates for outbursting YSOs. 72 of them have a $>5\sigma$
detection in the two WISE bands. As for the other samples, we checked the
Spitzer and WISE images for all these candidates. For the clear majority of them
(60/77) it turns out that they have neighbour star in 10" distance or less, 
which likely affects the WISE photometry. This indicates that the 6" criterion 
chosen in the preparation of the catalogue (see above) to account for the broad
WISE PSF was slightly too conservative. In addition, there are 7 objects within a
nebulosity. Again, this might cause problems in the WISE photometry. In all
these cases the image comparison excludes that the objects are in fact
significantly brighter in the WISE survey. The remaining 10 sources remain good
candidates and require further evaluation.

Glimpse provides for a fraction of the total area multiple epochs of Spitzer
photometry with baselines up to 1\,yr, particularly in the additional Glimpse-II
survey. Based on this information, the Glimpse catalogues exclude variable
sources for their final merged photometry. The Robitaille list, on the other
hand, uses the Glimpse-II first epoch photometry, and thus attempts to exclude
as few variables as possible \citep{2008AJ....136.2413R}, making this sample
suitable for our purposes.

In Appendix \ref{a1} we provide an estimate of the contamination by AGB stars
in the Robitaille catalogue, both in the entire sample and in the subsample
of variable sources. Among the variable candidates, the contamination is
negligible. The total sample of 7101 objects should contain about 3700-3800 YSOs;
the remaining sources are probably AGBs. 

\section{Follow-up observations}
\label{s3}

We summarise the results from the previous section and the selection of burst candidates in Table \ref{t10},
As outlined above, about 130 of the objects in the samples considered here show the signature of a 
brightness eruption when comparing Spitzer and WISE photometry (listed as 'highly variable' in Table 
\ref{t10}), but most of them are clearly spurious based on an inspection of the images. To verify our 
candidates, we re-observed a subset of them in the near-infrared. This was particularly important for 
objects which are confirmed to be brighter in WISE after visual inspection ('burst candidates'). If 
any of these sources is indeed a burst (as defined in Sect. \ref{s1}), we expect it to be several 
magnitudes brighter in the near-infrared compared with 2MASS. In total, we observed 20 from the highly
variable objects, including 13 out of 17 burst candidates. By design, these observations also
covered some of the spurious detections, to double-check our rejection based on visual examination.

\begin{table}
\caption{Summary of samples used in this paper
\label{t10}}
\begin{tabular}{ll}
\hline
Sample & No.\\
\hline
C2D total (Sect. \ref{c2d})          & 1478 \\
-- with WISE                         & 1301 \\
-- highly variable                   & 23   \\
-- burst candidates                   & 0 \\
\hline
Cluster total (Sect. \ref{sampleb})  & 2548 \\
-- with WISE                         & 1745 \\
-- highly variable                   & 24 \\
-- burst candidates                   & 5\\
\hline
Complementary (Sect. \ref{compl})    & \\
-- NGC2264 with WISE                 & 355 \\
-- Taurus with WISE                  & 320 \\
-- NaP with WISE                     & 935 \\
-- highly variable                   & 10 \\
-- burst candidates                   & 2 \\
\hline
Robitaille, (Sect. \ref{samplec})    & 18949 \\
-- cleaned, with WISE                & 7101 \\
-- highly variable                   & 77 \\
-- burst candidates                  & 10 \\
\hline
\end{tabular}
\end{table}

We used the 1.3\,m telescope at the Cerro Tololo International Observatory with the instrument Andicam, 
a double-channel camera which allows us to take optical and near-infrared images simultaneously. The 
follow-up observations were taken as part of the SMARTS collaboration in program DUBLIN-11B-001 and 
DUBLIN-12A-001 (PI: A. Scholz). For all objects we obtained optical images in the R- and I-bands 
($3 \times 120$\,sec exposures) and near-infrared images in the J- and either K- or H-band 
($5 \times 30$\,sec in a 5-position dither pattern), but only the near-infrared images are used 
here, since most objects are embedded and hence invisible in the optical. 

We carried out a standard image reduction, including sky subtraction and flatfielding, and aperture
photometry. The near-infrared photometry was calibrated in comparison with 2-5 other stars in the 
images, which are listed in the 2MASS point-source catalogue. For about half of the objects the new 
photometry is consistent with the 2MASS values, i.e. the variation in the mid-infrared cannot be caused
by a long-lasting eruptive event. Most of the remaining objects have only variations with $<1$\,mag, which is 
too little to qualify as an accretion burst according to our criterion (see Sect. \ref{s1}). We list these 
excluded objects in Table \ref{t2}. In particular, our follow-up observations confirmed that none 
of the highly variable objects seen as spurious in the visual inspection was misclassified.

\begin{table*}
\caption{Highly variable objects found in this study by comparing Spitzer and WISE photometry and ruled out 
by SMARTS photometry. The offsets between WISE and Spitzer photometry are listed in columns 4 and 5; 2MASS
photometry in columns 6 and 7. Our SMARTS photometry with the observing dates and the most likely reason for 
the photometry offset in the mid-infrared data are contained in columns 8 and 9.
\label{t2}}
\begin{tabular}{llccccclll}
\hline
$\alpha$ (J2000) & $\delta$ (J2000) & Sample & $\Delta$3.6$\,\mu m$ & $\Delta$4.5$\,\mu m$ & J2M & K2M & SMARTS photometry & Comments\\
\hline
15 42 20.96   & -52 48 46.4 & A & 1.744 & 1.732 & 13.879 &  7.944 & J=13.9, K=8.0 (12-07-18)  & \\ 
15 42 31.06   & -52 47 16.9 & A & 2.814 & 1.250 &  9.636 &  7.390 & J=9.6, K=7.4 (12-07-18)   & saturation in Spitzer\\
16 31 33.84   & -24 04 46.8 & A & 1.010 & 1.624 & 12.526 & 10.049 & J=12.9, K=9.5 (12-05-02)  & saturation in Spitzer\\ 
18 29 01.76   &  00 29 47.3 & A & 1.210 & 2.393 & 11.170 &  9.215 & J=11.2, K=9.2 (12-05-08)  & companion in WISE\\
20 06 57.23   &	 27 26 35.8 & A & 1.171	& 2.219 & -      & -	  & J$>$16.5, K$>$15 (12-09-09) & \\
06 07 29.50   & -06 24 45.3 & A & 1.094 & 1.639 & -      & 13.962 & J$>$16.5, K=13.4 (12-09-11) & variable$^1$\\
\hline
14 32 27.34   &	-60 56 26.9 & B & 1.685 & 1.524 & 14.695 & 12.263 & J=15.0, K=12.5 (12-06-04) & bright neighbour\\
16 02 24.58   & -51 28 47.7 & B & 1.854 & 1.924 & 14.922 & 12.439 & J=14.7, K=12.1 (12-06-07) & bright neighbour$^2$\\
14 50 13.85   & -59 03 54.8 & B & 2.272 & 2.583 & 13.433 & 12.035 & J=13.4, K=11.9 (12-06-22) & bright neighbour\\
17 22 21.08   & -37 33 29.1 & B & 1.690 & 1.782 & 14.103 & 12.304 & J=14.5, K=12.4 (12-06-24) & bright neighbour\\ 
16 17 36.21   & -50 56 01.8 & B & 1.136 & 1.106 & -      & 10.974 & J$>$16.5, K=11.4 (12-09-10) & neighbours\\
16 38 15.17   & -47 48 24.4 & B & 1.113 & 1.593 & -      & 11.140 & J$>$16.5, K=13.1 (12-07-22) & variable$^1$\\
18 15 13.88   & -17 21 06.0 & B & 1.078 & 1.407 & 12.488 & 11.625 & J=12.5, H=12.1 (12-08-05) & \\
18 18 18.42   & -16 27 09.8 & B & 1.437 & 1.412 & -      & 11.909 & J$>$16.5, H$>$15.0 (12-08-07)& \\
18 53 58.11   & +01 43 44.4 & B & 1.240 & 1.655 & -      & 12.702 & J$>$16.5, K=13.5 (12-08-19)& variable$^1$\\
18 57 47.12   & +03 30 16.9 & B & 1.232 & 1.457 & -      & 13.393 & J$>$15, K$>$12 (12-08-20) & affected by clouds\\
19 23 52.82   & +14 38 03.5 & B & 1.484 & 1.510 & -      & 13.814 & J=15.6, K=12.9 (12-08-24) & neighbours \\
\hline
\end{tabular}

$^1$ variability does not match the typical signature of an accretion burst\\ 
$^2$ object identification ambiguous, two sources in aperture
\end{table*}

\begin{table*}
\caption{Burst candidates found in this study by comparing Spitzer and WISE photometry. The primary criterion
is a brightness increase by 1\,mag in the two channels at 3.6 and 4.5$\,\mu m$.
\label{t1}}
\begin{tabular}{llcccccl}
\hline
$\alpha$ (J2000) & $\delta$ (J2000) & Sample & $\Delta$3.6$\,\mu m$ & $\Delta$4.5$\,\mu m$ & J2M & K2M & Comments\\
\hline
\hline
03 27 05.84  & +58 43 47.8 & A & 1.458 & 1.507 & -      & 14.846 & in AFGL490\\
16 26 36.82  & -24 19 00.3 & A & 1.526 & 1.631 & 16.823 & 14.756 & ISO-Oph-50, Sect. \ref{comm}\\
20 50 09.40  & +44 26 52.2 & A & 1.698 & 1.995 & -      & 14.455 & Northamerica/Pelican\\ 
20 51 26.23  & +44 05 23.9 & A & 2.911 & 2.903 & -	& -	 & V2492 Cyg, Sect. \ref{comm}\\
22 19 32.95  & +63 33 16.2 & A & 1.857 & 2.019 & 16.102 & -      & in S140-North\\
\hline    
16 44 37.21  & -46 04 01.1 & B & 1.129 & 1.397 & -      & 13.750 & 2M1644-4604, Sect. \ref{comm}\\
15 11 13.68  & -59 02 36.1 & B & 1.050 & 1.289 & -      & 12.541 & 2M1511-5902, Sect. \ref{comm}\\
\hline
\end{tabular}
\end{table*}

From the 17 burst candidates, we observed 13 and rejected 10 of them (contained in Table \ref{t2}). The remaining
7 are listed in Table \ref{t1}. Two objects not previously known are confirmed by our SMARTS photometry as eruptive 
variables and are good burst candidates: 2MASS J16443712-4604017 (hereafter 2M1644-4604) and 2MASS J15111357-5902366 
(hereafter 2M1511-5902). These two objects, together with the two previously identified possible burst objects ISO-Oph-50
and V2592 Cyg, are discussed in more detail in Sect. \ref{comm}. Three objects remain unconfirmed because they are too 
far north to be observed from Cerro Tololo. Given the fact that most of our candidates so far have been ruled out by 
follow-up observations, the likelihood that one of these three turns out to be bursts is fairly low.

\subsection{Comments on specific objects}
\label{comm}

{\bf ISO-Oph-50:} As pointed out in Sect. \ref{sampleb}, one of the candidates from the Cluster sample, ISO-Oph-50 (or 
CFHTWIR-Oph 30) was previously suspected to be an outbursting YSO \citep{2008A&A...485..155A}, maybe of EXor type. 
\citet{2012A&A...539A.151A} measure an optical spectral type of M3.25 for this object. In Table \ref{t3} we list 
the available photometry in the H-band (the band with the most measurements) for this object, including a new value 
obtained from our SMARTS imaging on August 8 2012. Out of 6 epochs, 3 are around $H=14$\,mag, while the others are 
around $H=16$\,mag. In addition, there is evidence for significant variability on short timescales of days and weeks
\citep{2008A&A...485..155A}. This behaviour is not comparable to typical stars undergoing accretion-related eruptions. 
EX Lupi, probably the best studied YSO with short-term and recurring accretion bursts of EXor-type, had 4 bursts in 9 
years between 1995 and 2004, but all four were different in amplitudes. Taken together, the bursts lasted in total 
about 1 year, i.e. $\sim 10$\% of the entire time \citep{2007AJ....133.2679H}. ISO-Oph-50 is much more often found 
in the bright state. Also, as noted by \citet{2012A&A...539A.151A}, the object becomes bluer when fainter, which
is not typical for accretion-related bursts. It can safely be concluded that this source is not an accretion burst, 
in particular not a FU Ori object. 
 
Apart from the variability, the most remarkable feature of ISO-Oph-50 is its low luminosity. At the age and distance of the
$\rho$-Oph star forming region, a M3 star would be expected to have an H-band magnitude of 8-10, i.e. even
with $A_V =10$\,mag it would be brighter than $H=12$\,mag, whereas the object is never observed to be brighter than
$H=13$\,mag. The luminosity of this source, estimated from the J-band magnitude, is  $\log{(L/L_{\odot})} \sim -2.56$ 
(Alves de Oliveira, priv. comm), which is more than two orders of magnitude too low for this spectral type. 
Given that and the colour trend in the variability, the variability is likely related to the disk. We speculate that 
the most likely cause for the variations is an rotating, inhomogenuous edge-on disk. \citet{2012A&A...539A.151A} 
come to a similar conclusion, but also invoke the presence of a (hypothetical) companion to explain the variations.
Monitoring with simultaneous measurements in multiple bands and detailed modeling is needed to constrain the nature
of this source.

\begin{table}
\caption{H-band photometry for ISO-Oph-50
\label{t3}}
\begin{tabular}{lll}
\hline
Epoch & H (mag) & Comments \\
\hline
1993-94        & 13.93          & \citet{1997ApJS..112..109B} \\
Apr 1999       & 16.01          & 2MASS\\
Apr 2005       & 15.91          & UKIDSS/GCS\\
May 2005       & 15.9           & Alves de Oliveira et al. (2008)\\
Jun 2006       & 13.3-14.7      & Alves de Oliveira et al. (2008)\\
Aug 2012       & 14.1           & SMARTS (also, $J\sim 16.4$\,mag)\\
\hline
\end{tabular}
\end{table}

{\bf V2492 Cyg:} This object was already known in the literature as an outbursting protostar although 
it does not fit into the FU Ori category \citep{2011A&A...527A.133K,2011AJ....141...40C}. It was confirmed by 
our Spitzer-WISE comparison. Its magnitude differences in the mid-infrared are almost 3\,mag and very large 
compared with most of our other candidates. In optical and near-infrared bands \citet{2011A&A...527A.133K} report 
amplitudes of more than 5\,mag. 

\begin{figure*}
\includegraphics[width=8.6cm]{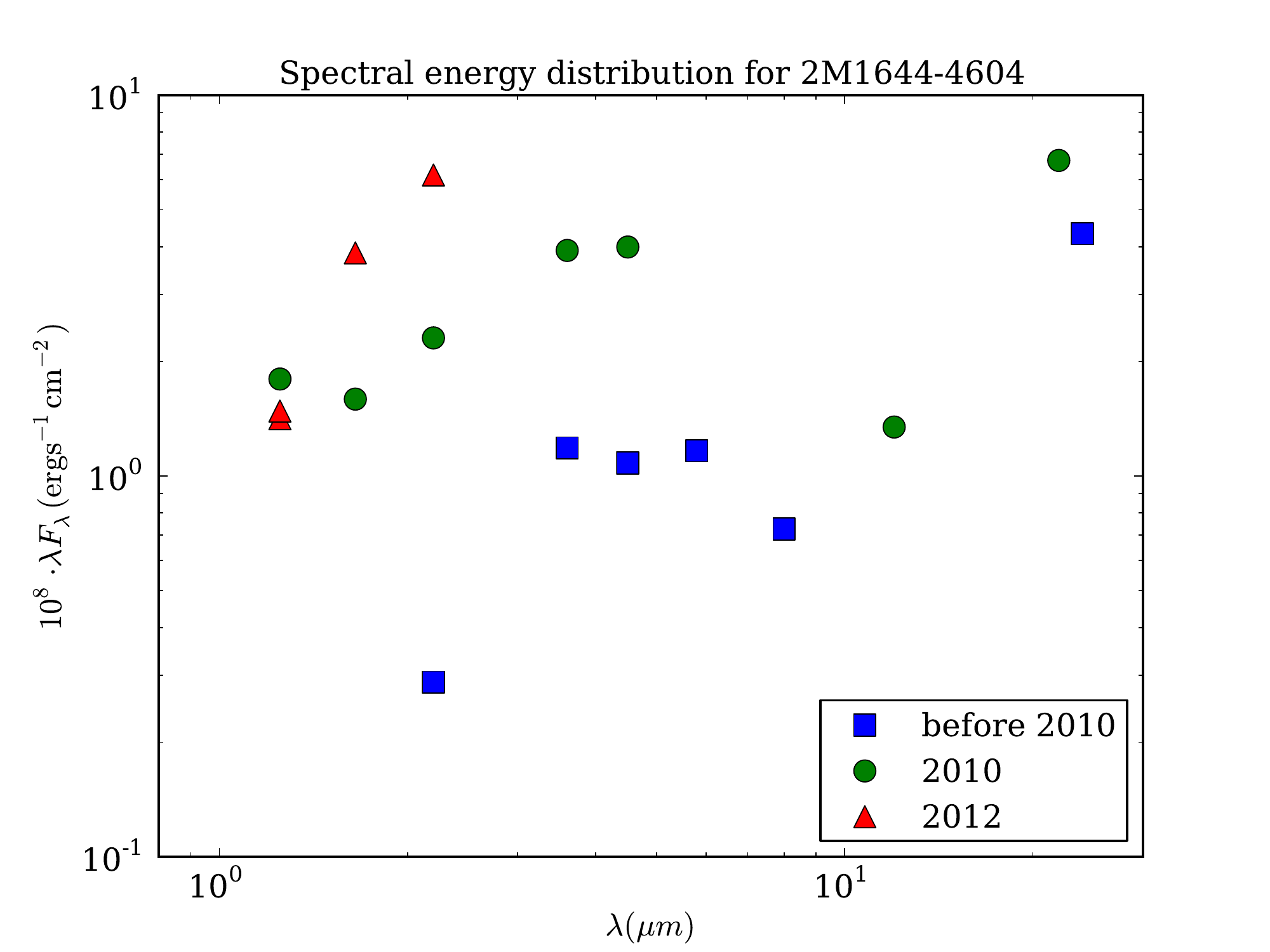}
\includegraphics[width=8.6cm]{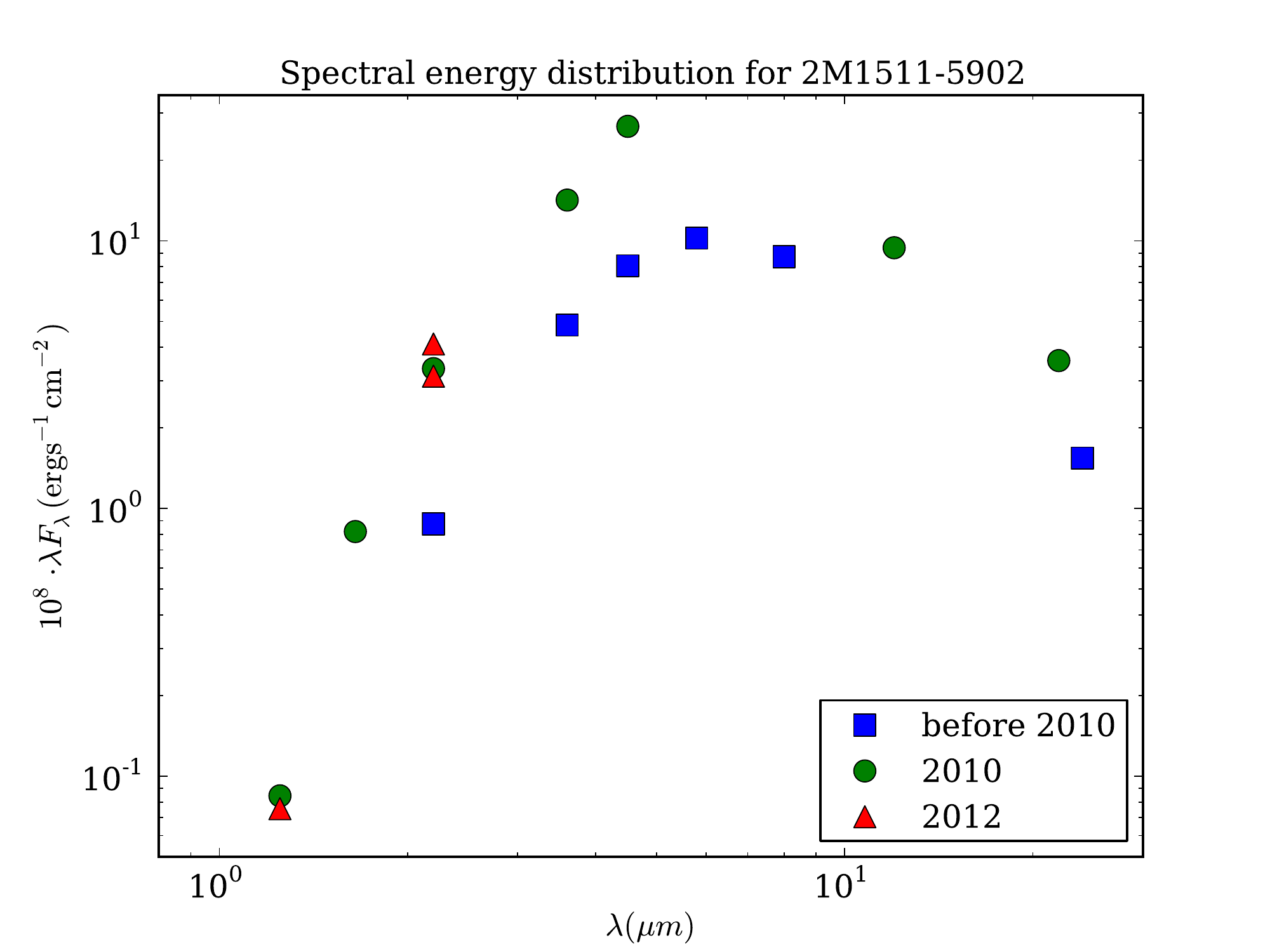}
\caption{\label{sed} Spectral energy distributions for the two newly identified eruptive YSO candidates,
2M1644-4604 (left panel) and 2M1511-5902 (right panel). Blue datapoints show photometry pre-2010 from
2MASS and Spitzer. Green datapoints show the 2010 data from VVV and WISE. Red datapoints are from our own
observations in 2012. The photometric errorbars are $<10\%$ and thus small compared with the size of the
symbols. Note that the datapoints plotted in blue are {\it not} from the same year (see Tables \ref{t4} and
\ref{t5} for the complete list of epochs).}
\end{figure*}

{\bf 2M1644-4604:} As pointed out above, this object was identified as a new eruptive variable and possible accretion 
burst. The available photometry for the object is summarised in Table \ref{t4}, including near-infrared data from the 
first data release from
the VISTA/VVV survey \citep{2012A&A...537A.107S}. In Fig. \ref{sed} we show the spectral energy distribution pre- and 
post-burst, including our new datapoints from 2012. In near-infrared data from 2010-12 the source is much brighter than 
in 2MASS -- more than 4\,mag in J, more than 3\,mag in H, and more than 2\,mag in K. The near-infrared photometry 
indicates significant evolution from 2010 to 2012. In addition, the object has become more than 1\,mag brighter at 3.6 
and 4.5$\,\mu m$. The WISE flux at $22\,\mu m$ is slightly brighter than the 24$\,\mu m$ from Spitzer as well. The 
difference in magnitudes is increasing towards shorter wavelengths, i.e. the object became bluer during the burst. 
The near-infrared photometry indicates a position below the reddening path, i.e. it is indeed a likely YSO (see Appendix 
\ref{a1}). Spectroscopic follow-up observations are in preparation, to confirm its youth and to look for evidence of 
enhanced accretion.

\begin{table}
\caption{Photometry for 2M1644-4604
\label{t4}}
\begin{tabular}{llll}
\hline
Epoch & Band & Magnitude & Comments \\
\hline
1999-05-20  & J    & $>$17.34  & 2MASS\\
1999-05-20  & H    & $>$16.06  & 2MASS\\
1999-05-20  & K$_s$& 13.75     & 2MASS\\
2010-05-09  & J    & 13.32     & VVV\\
2010-05-09  & H    & 12.67     & VVV\\
2010-08-18  & K$_s$& 11.49     & VVV\\
2012-07-28  & J    & 13.58     & SMARTS\\
2012-07-28  & H    & 11.71     & SMARTS\\
2012-09-15  & J    & 13.53     & SMARTS\\
2012-09-15  & K$_s$& 10.42     & SMARTS\\
\hline      
2004-09-05    & 3.6$\,\mu m$ & 10.74 & Glimpse\\
2004-09-05    & 4.5$\,\mu m$ & 10.11 & Glimpse\\
2004-09-05    & 5.8$\,\mu m$ &  9.27 & Glimpse\\
2004-09-05    & 8.0$\,\mu m$ &  8.80 & Glimpse\\	  
2006-10-03    & 24$\,\mu m$  &  3.29 & \citet{2008AJ....136.2413R}\\
2010-06-02$^1$  & 3.6$\,\mu m$ & 9.548 & WISE\\	  
2010-06-02$^1$  & 4.5$\,\mu m$ & 8.644 & WISE\\      
2010-06-02$^1$  & 12$\,\mu m$  & 6.925 & WISE\\	  
2010-06-02$^1$  & 22$\,\mu m$  & 3.071 & WISE\\	        
\hline
\end{tabular}

$^1$ Several epochs from 2010-06-02 to 2010-06-05
\end{table}

{\bf 2M1511-5902:} This is the second possible new accretion burst identified in our survey. We summarise
the available photometry, including our own follow-up, in Table \ref{t5}. The spectral energy distribution
is plotted in Fig. \ref{sed}. Comparing pre-2010 with 2010 datapoints, the object is 1.5\,mag brighter in 
K-band, 1.1\,mag at 3.6$\,\mu m$ and 1.3\,mag at 4.5$\,\mu m$. This trend is also seen at 22-24$\,\mu m$. 
Between 2010 and 2012 the changes are marginal, i.e. the brightening appears to be persistent. Similar to 
2M1644-4604, the colours indicate that this is indeed a YSO, but more follow-up observations are needed 
to confirm the nature of the source and to make sure that the brightening is indeed due to enhanced accretion.

\begin{table}
\caption{Photometry for 2M1511-5902
\label{t5}}
\begin{tabular}{llll}
\hline
Epoch & Band & Magnitude & Comments \\
\hline
1999-07-07  & J    & $>$17.59  & 2MASS\\
1999-05-07  & H    & $>$16.03  & 2MASS\\
1999-07-07  & K$_s$& 12.54     & 2MASS\\
2010-04-11  & J    & 16.64     & VVV\\
2010-04-11  & H    & 13.39     & VVV\\
2010-08-14  & K$_s$& 11.09     & VVV\\
2012-08-14  & J    & 16.76     & SMARTS\\
2012-08-14  & K$_s$& 10.86     & SMARTS\\
2012-09-12  & K$_s$& 11.16     & SMARTS\\
\hline      
2004-03-12    & 3.6$\,\mu m$ & 9.21 & Glimpse\\
2004-03-12    & 4.5$\,\mu m$ & 7.93 & Glimpse\\
2004-03-12    & 5.8$\,\mu m$ & 6.91 & Glimpse\\
2004-03-12    & 8.0$\,\mu m$ & 6.10 & Glimpse\\	  
2006-04-11    & 24$\,\mu m$  & 4.41 & \citet{2008AJ....136.2413R}\\
2010-02-21$^1$  & 3.6$\,\mu m$ & 8.148 & WISE\\	  
2010-02-21$^1$  & 4.5$\,\mu m$ & 6.578 & WISE\\      
2010-02-21$^1$  & 12$\,\mu m$  & 4.811 & WISE\\	  
2010-02-21$^1$  & 22$\,\mu m$  & 3.761 & WISE\\	        
\hline
\end{tabular}

$^1$ Several epochs from 2010-02-21 to 2010-02-23
\end{table}

\section{The statistics of accretion bursts}
\label{s4}

In this paper we have systematically searched for eruptive variables that may be accretion bursts fulfilling 
specific conditions outlined in Sect. \ref{model}. We find 1 known accretion burst and three more possible 
bursts in sample A and 2 probable bursts in sample B. In the following sections we will use this result to
derive constraints on the typical interval between bursts and compare with other constraints from theory and
observations. We will treat sample A and B separately, because they are significantly different
in terms of the typical ranges of stellar masses -- while sample A is dominated by low-mass stars with
masses around or below 1$\,M_{\odot}$, objects in sample B are much further away and will therefore 
have on average masses higher than 1$\,M_{\odot}$.

\subsection{Statistical estimate of the burst frequency}
\label{stat}


For one burst out of 4000 stars and an epoch difference of 5\,yr a crude estimate following
the arguments given in Sect. \ref{model} gives a burst interval of 20000\,yr. To obtain confidence
intervals for this number, we implemented simple Monte-Carlo simulations: For a given burst interval, 
we calculated the probability that a star experiences a burst over a given epoch difference. 
For each star we then obtain a random number between 0 and 1 and count the ones for which this 
number exceeds the burst probability. This procedure was repeated over 10000 runs; then we can 
calculate the probability to find a given number of bursts (in our case one).

\begin{figure}
\includegraphics[width=8cm,angle=0]{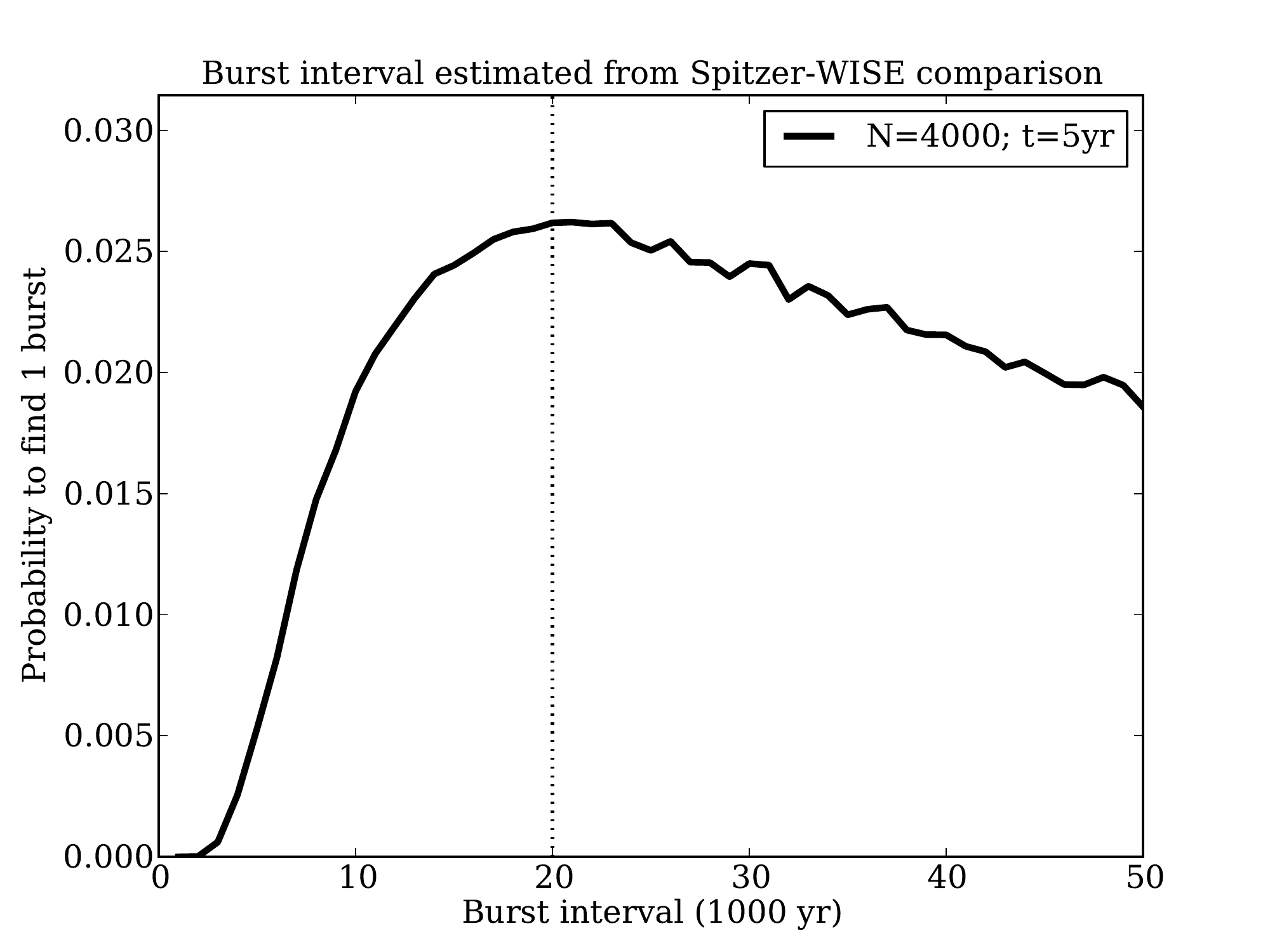}
\caption{Monte-Carlo simulations of burst statistics for the Spitzer-WISE comparison. Probability to 
find a 1 burst as a function of burst interval for a total sample of 4000 stars and a epoch 
difference of 5\,yr. The 95\% lower limit is indicated by the dotted line.
\label{dc}}
\end{figure}

In Fig. \ref{dc} we show the results from this simulation when applied to the
Spitzer-WISE comparison. For an epoch difference of 5\,yr and a sample size of 4000 stars, the 
detection of one burst implies that we can rule out a burst interval below 20000\,yr with 95\% confidence.
The upper limit is not well-defined due to the poor statistics. For two bursts, the 95\% lower 
limit drops to around 10000\,yr, for 4 bursts to 3000\,yr. As noted above, 4 bursts is the most
conservative upper limit we derive from our survey. Thus, from the Spitzer-WISE comparison we can 
derive a robust lower limit for the burst interval in the range of $10^4$\,yr. 

A similar type of simulation was used to derive an estimate of the burst frequency from the known
FU Ori outbursts. Among the known FU Ori objects, 10 have an observed burst event, 9 of them between
1936 and 1999, the 10th probably before 1888 (Reipurth \& Aspin 2010). Since most of these objects 
have been found serendipitously and outside systematic surveys, the choice of parameters (number 
of monitored stars $N$ and time baseline $t$) for the simulation is not trivial. For a rough estimate 
we assume that optical surveys based on photographic observations had access to at most about 1000 
young stars in the solar neighbourhood. We note that a few more possible FU Ori outbursts have been 
found over the past 3 years \citep{2011ApJ...730...80M,2012ApJ...748L...5R,2011A&A...526L...1C}.

\begin{figure}
\includegraphics[width=8cm,angle=0]{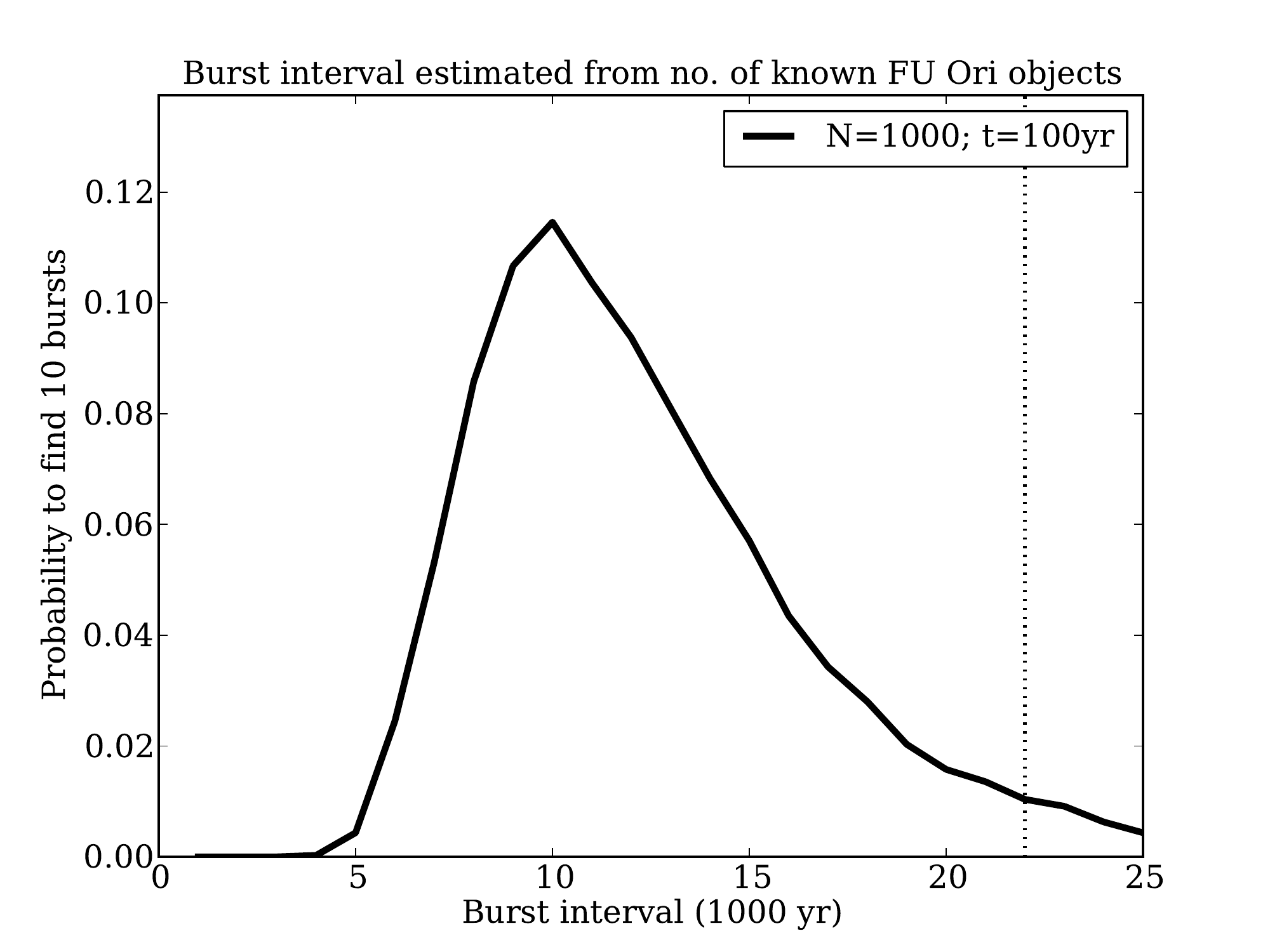}
\caption{Monte-Carlo simulations of burst statistics for the known FU Ori events. Probability to find 10
bursts as a function of burst interval assuming a total sample size of 1000 and an epoch difference of 100\,yr.
The 95\% upper limit is indicated by the dotted line.
\label{dcfuori}}
\end{figure}

In Fig. \ref{dcfuori} we show the probability to find 10 bursts as a function
of interval. For $N=1000$ and $t=100$\,yr the burst interval is in the range of 
8000-12000\,yr, with an upper limit at 22000\,yr (95\% confidence) and a lower limit around 5000\,yr. 
Using $t=50$\,yr (maybe more plausible, given that only 2 events have been recorded prior to 1940) these 
numbers would be halved. On the other hand, doubling the sample size to 2000 stars would also double the 
estimated interval. Given the uncertainties in the choice of the parameters, we conclude that the known 
FU Ori events constrain the burst interval to 2000-50000\,yr.

Taken these numbers together, the interval between consecutive accretion bursts
with a) a mass accretion rate increasing to $10^{-6}\,M_{\odot}$yr$^{-1}$ or more, b) a rise time of 
$<5$\,yr and c) a decline time of $>5$\,yr is most likely in the range of $10^4$\,yr, and with high 
confidence between 5000 and 50000\,yr. We note that this is consistent with lower limits derived 
from near-infrared surveys of YSOs, which give intervals longer than 2000-3000\,yr 
\citep{2001AJ....121.3160C,2012MNRAS.420.1495S}. 

\subsection{Comparison with constraints from outflows}

FU Ori bursts are associated with strongly enhanced rates of mass accretion
as well as enhanced mass outflow rates \citep[e.g.][]{1996ARA&A..34..207H}. The
properties of jets and outflows may therefore be connected with these events
\citep{1985A&A...143..435R}. If this is the case, then the burst interval
should be reflected in a corresponding time scale for jets and outflows. One
possibility is that accretion bursts also power enhanced collimated ejection. 
This in turn would lead to the formation of new jet knots with separations of the 
order of the burst frequency. A second option is that accretion bursts either cause
an enhanced mass outflow rate and thus trigger strong outflow activity, or 
destabilise the large-scale magnetic field and thus terminate an episode
of collimated outflow activity, i.e. switch to a wide-angled wind.

Over the last decade the census of jets and outflows in nearby star forming
regions has become more and more complete \citep{2007prpl.conf..215B}. Furthermore, 
there are now large-scale unbiased surveys to establish outflow properties along the 
Galactic Plane \citep[e.g.][]{2011MNRAS.413..480F}. \citet{2012arXiv1206.5095I} have 
recently determined typical dynamical jet lifetimes and time gaps between emission knots 
for an unbiased sample of 130 jets and outflows from the Galactic Plane survey UWISH2. 
They find that the time gaps between emission knots are of the order of 10$^3$\,yrs and 
the dynamical lifetimes are an order of magnitude larger, i.e. 10$^4$\,yrs. Interestingly, 
this is in line with earlier estimates obtained for the molecular outflows of FU Ori stars 
\citep{1994ApJ...424..793E}.

As discussed in Sect. \ref{stat} our statistical limits on the burst interval from 
the Spitzer-WISE comparison as well as the constraint obtained from the known sample
of FU Oris gives $\sim 10^4$\,yr for the burst frequency. Values on the order of $10^3$\,yr 
are highly unlikely. Thus, only the dynamical timescales of outflows can be identified 
with the timescale between consecutive accretion bursts, not the separation of outflow knots. 
This could mean that a strong burst triggers the formation of a new outflow or terminates
the collimated outflow activity. The outflow knots then represent variations 
on shorter timescales.

One possible caveat in this comparison is that the large-scale outflows are mostly driven 
by sources in an early evolutionary stage whereas our analysis is biased towards Class II 
sources. It remains to be explored whether the frequency of eruptive variables increases
significantly in the Class I stage.

\subsection{Comparison with constraints from protostellar luminosities}

The statistics of the protostellar luminosities does not provide a direct constraint
on the burst frequency, as defined in this paper, but can be used to estimate the duty cycle. 
The best observational limit for this number comes from the Spitzer C2D 
survey. \citet{2009ApJS..181..321E} estimate that for a specific model stars accrete
half of their mass in $\sim 40000$\,yr, which corresponds to 7\% of the Class I lifetime of 
$\sim 0.5$\,Myr. On the other hand, by comparing the C2D dataset with models for episodic 
accretion driven by gravitational instabilities, \citet{2012ApJ...747...52D} find that YSOs 
spend on average only 1.3\% of their total time of $\sim 1$\,Myr in accretion bursts (0-12\%), 
i.e. around 13000\,yr. 

The protostellar lifetimes in these estimates (0.5 and 1.0\,Myr) are comparable to the typical 
ages of the clusters and star forming regions covered in our analysis, i.e. a comparison with 
our results is valid. For that purpose, however, we need to assume a typical
duration for the bursts. Assuming that the bursts occur over 0.5\,Myr the burst interval of 
$10^4$\,yr is consistent with a duty cycle of 7\% if the burst duration is on the order of 
800\,yr. For 1\,Myr and a 1.3\% duty cycle, the burst duration has to be 130\,yr to be 
consistent with our interval. These values are plausible given the slow decline observed in 
the most extreme known FU Ori bursts. Thus, assuming burst durations of hundreds of years
our constraint is consistent with the ones derived from protostellar luminosities.

\subsection{Further discussion}

While our estimate is robust for the assumptions given in Sect. \ref{s1}, two additional
caveats should be kept in mind when interpreting our findings. First, it is conceivable
that episodic accretion does not affect all stars in the same way. Some of the bursts 
could be triggered by mechanisms that are not applicable to all known YSOs, for example, 
the presence of a companion or disk-planet interaction (see Sect. \ref{s0}). This would 
imply that the frequency of bursts is strongly variable among protostars and it is not
valid to extrapolate from the sample of known bursts. 

Second, accretion bursts are often thought to occur mostly in the Class I stage of the
protostellar evolution and less frequent at Class II stage. This is supported by the
finding that the known FU Ori-type objects tend to be more comparable to Class I objects in 
terms of their disk/envelope properties \citep{2001ApJS..134..115S}. Our samples include 
objects in these early stages -- probably around one quarter to one third -- but they are 
still dominated by the slightly older Class II objects. If we would limit the statistical
analysis to the Class I objects, the burst interval could be by a factor of 3-4 lower than 
in our estimate. This, however, would conflict with the constraint from the known FU Oris. 
Therefore, we do not think that the burst interval will be significantly below 5000\,yr, 
even if bursts only occur in the Class I stage.

Constraining the burst interval for a simple model of accretion bursts as shown in Sect. 
\ref{model} is only the first step in a characterisation of the accretion history of YSOs. 
Various arguments suggest that the accretion history is in fact more complex 
than the simple model that is tested and constrained here 
\citep[e.g.][]{2011ApJ...736...53O,2012ApJ...747...52D}. Indeed, as already acknowledged in
Sect. \ref{model}, the known accretion bursts show a significant degree of diversity in rise
time, decline time, and amplitude. Thus, the long-term goal should be to derive the frequency 
spectrum of bursts, and not only the interval. In addition, episodic accretion events may be 
combined with more gradual trends in the mass accretion rate that cannot be captured on 
timescales of years. 

With only few epochs of photometry available for most of the YSOs in the solar neighbourhood,
deriving direct observational constraints for these more complex scenarios is not feasible at the 
moment. With our approach, we only probe the contrast between the strong bursts and the quiescent phases. 
Substantial accretion rate variations in the quiescent phases would mask the signals and prevent 
a detection. Long-term monitoring of large samples or follow-up on the variable objects below our 
threshold of 1.0\,mag will yield more information about the presence and characteristic of additional 
variations in the accretion histories of YSOs. The observational record of accretion histories
will become more complete with new time-domain surveys like Pan-Starrs, VISTA/VVV, Gaia, and ultimately
LSST. Out of these four, however, only VVV operates in the infrared and has access to the embedded, 
strongly reddened populations of YSOs. 

\section{Conclusions}

We have searched for eruptive variables among YSOs by comparing Spitzer and WISE photometry. 
In our first sample of $\sim 4000$ nearby YSOs, we find one previously known outbursting protostar
and three more possible variables with an eruption of $>1$\,mag at 3.6 and 4.5$\,\mu m$. In a
second sample of $\sim 4000$\,YSOs in the Galactic plane we find two new eruptive variables which
may be outbursting protostars. Based on the statistics of these findings, we estimate that
long-lasting, strong accretion bursts in protostars occur with intervals of $\sim 10^4$\,yr, with
high confidence between 5000 and 50000\,yr. For this estimate we assume that additional variability
is small compared with these events and that episodic accretion affects all stars in the same way.
The estimate is consistent with constraints from protostellar luminosities. It is also comparable to 
the dynamical timescales of protostellar outflows, indicating that accretion bursts may be responsible
for either triggering or terminating large-scale outflows.

\section*{Acknowledgments}
This publication makes use of data products from the Wide-field Infrared Survey Explorer, 
which is a joint project of the University of California, Los Angeles, and the Jet Propulsion 
Laboratory/California Institute of Technology, funded by the National Aeronautics and Space 
Administration. We also use data products from the Two Micron All Sky Survey, which is a joint 
project of the University of Massachusetts and the Infrared Processing and Analysis Center/California 
Institute of Technology, funded by the National Aeronautics and Space Administration and the 
National Science Foundation
This work also makes use of observations made with the Spitzer Space Telescope, 
which is operated by the Jet Propulsion Laboratory, California Institute of Technology under a 
contract with NASA. 
Part of this work was funded by the Science Foundation Ireland through grant 
no. 10/RFP/AST2780.

\appendix

\section{The Robitaille catalogue -- separation of YSOs and AGB stars}
\label{a1}

\begin{figure}
\includegraphics[width=6.6cm,angle=-90]{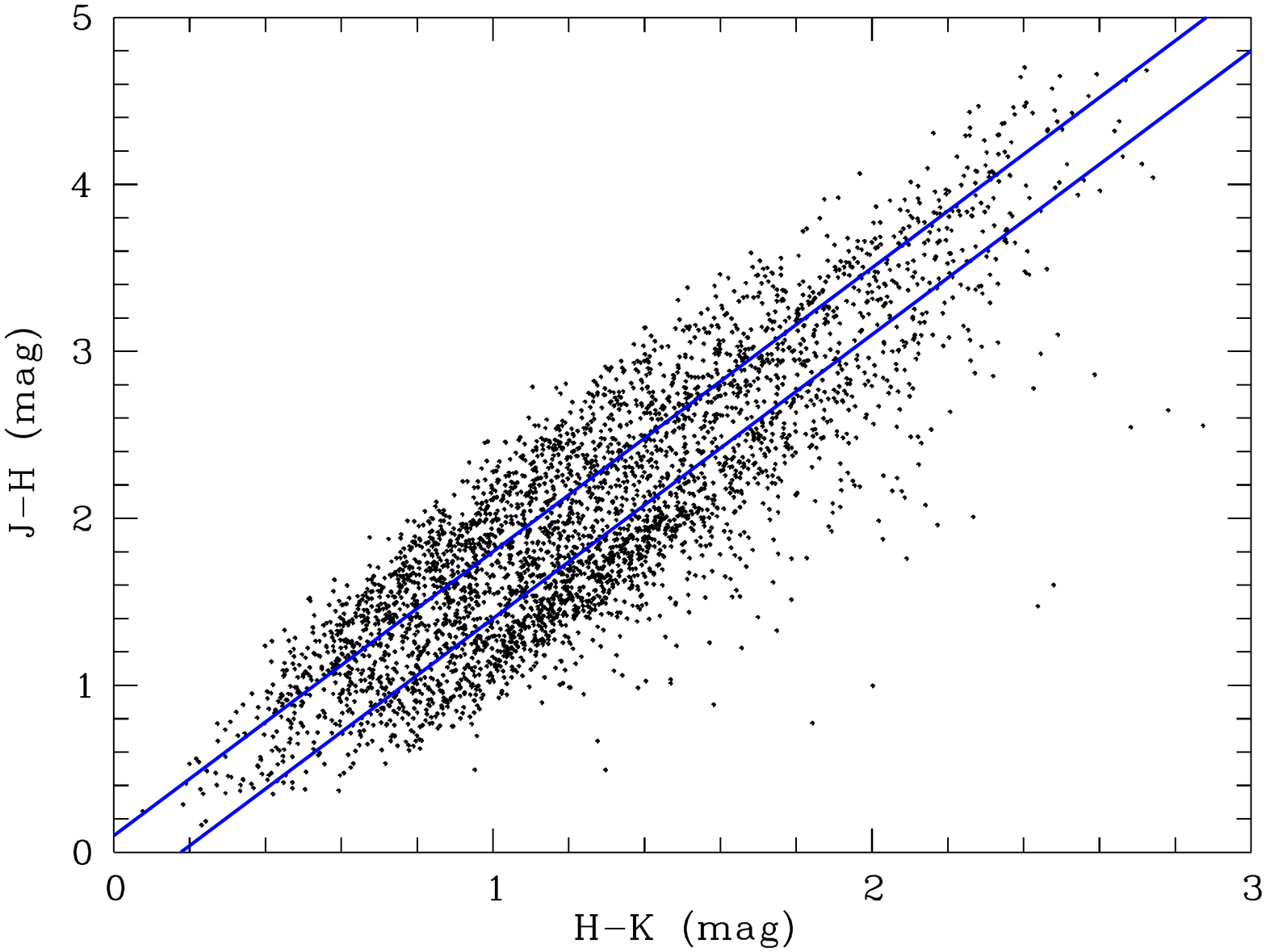}
\caption{\label{jhvshk} Near-infrared J-H vs H-K colour-colour diagram of the 3709 objects
in the Robitaille sample with JHK detection. The two solid lines indicate the two seperators
used to identify YSOs (below bottom line) and AGB stars (above top line). See text for
more information.}
\end{figure}

In order to estimate the fraction of AGB stars in the Robitaille sample, we started 
with our full sample of 7101 objects as defined in Sect. \ref{samplec}, and selected 
all objects with near-infrared colours from 2MASS in each band. Using the (H-K, J-H)
colour-colour diagram for this subsample of 3709 objects, we can establish three
distinct groups (see Fig.\,\ref{jhvshk}). Objects below the 'standard' reddening
band (defined as J-H $<$ -0.3 + 1.7 $\times$ H-K, or {\it seperator1}) are clearly
YSOs with K-band excess emission (YSO1a, hereafter -- 1389 objects). The group
at the top of the reddening band (J-H $>$ 0.1 + 1.7 $\times$ H-K, or {\it seperator2})
are clearly AGB objects (AGB1a, hereafter -- 1269 objects). The objects
inbetween are probably a mix of AGB stars and YSOs (YSO2a and AGB2a, hereafter
-- 1051 objects).

\begin{figure}
\includegraphics[width=6.6cm,angle=-90]{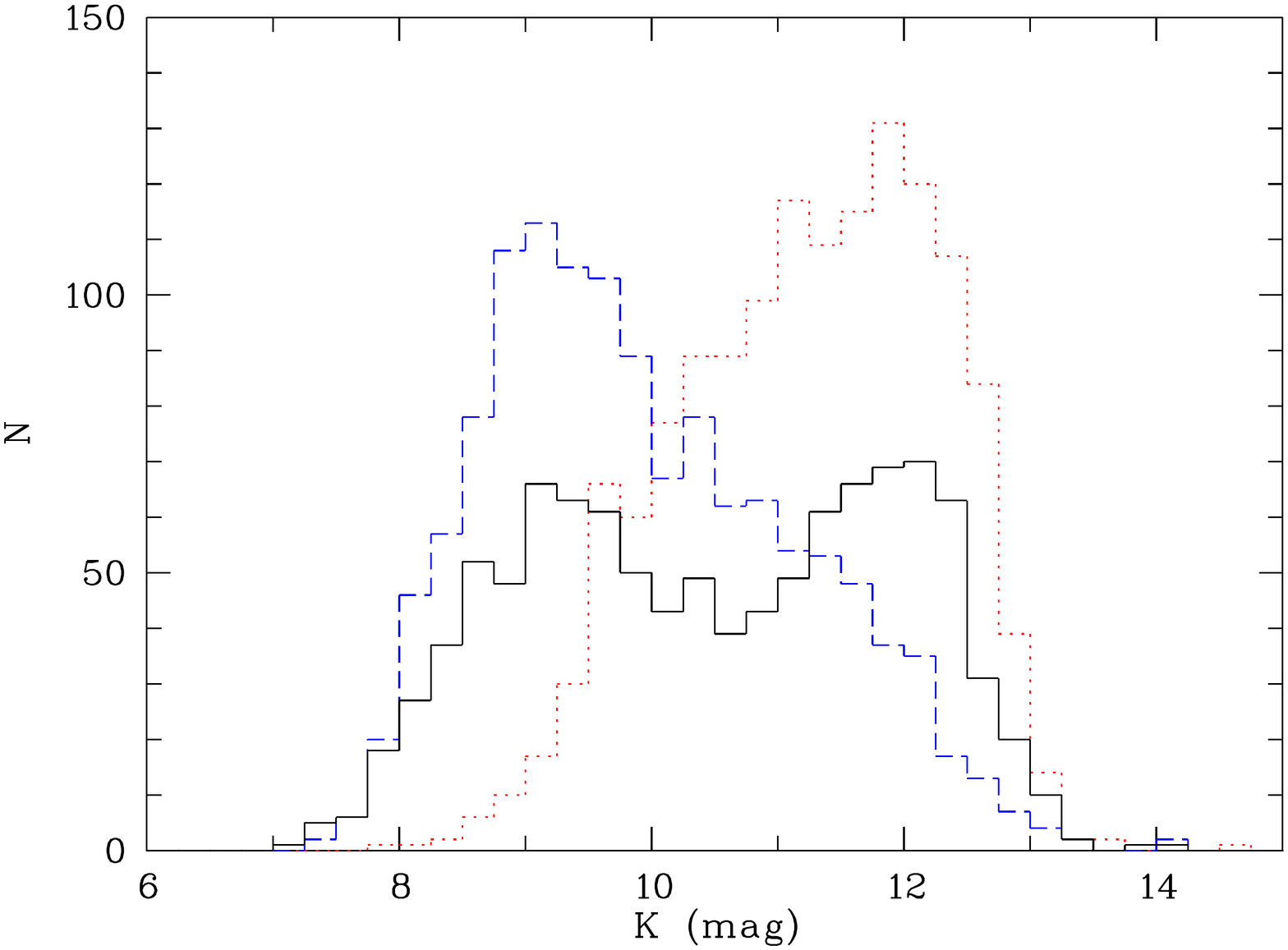}
\caption{\label{khist} K-band histogram of the JHK detected objects in the Robitaille
sample. The red dotted line is the YSO1 sample, the blue dashed line the AGB1 sample
and the solid black line the YSO2+AGB2 sample.}
\end{figure}

We investigate if there is a further way to separate these sources by plotting a
K-band luminosity function of YSO1a and AGB1a (see Fig.\,\ref{khist}). While
YSO1a shows a well defined peak between 11 and 13\,mag in K, the AGB1a sample is
brighter, mostly confined to 8-10\,mag with a tail of fainter objects. The
K-band luminosity function of AGB2a and YSO2a together (also shown in
Fig.\,\ref{khist}), is a clear mix of the the other two, with two peaks. Based
on this plot, we defined four 'clean' samples of objects:

\begin{enumerate}
\item YSO1: below {\it seperator1}, 11$<$K$<$13 (819 objects)
\item AGB1: above {\it seperator2}, 8$<$K$<$10 (696 objects)
\item YSO2: between {\it seperator1} and {\it seperator2}, 11$<$K$<$13 (428 objects)
\item AGB2: between {\it seperator1} and {\it seperator2}, 8$<$K$<$10 (399 objects)
\end{enumerate}

These objects are taken out of all 3709 objects which have a JHK detektion as
well as no 'bright' neighbours in Glimpse as discussed above (Sect. \ref{samplec}). 
With the described magnitude cuts there are 2342 objects remaining, from which 1247 
(53\,\%) are YSOs and 1095 (47\,\%) are AGB stars, in line with the percentages given 
in \citet{2008AJ....136.2413R}. We obtain a very similar result using only the YSO1a 
and AGB1a samples without K-band magnitude cuts (1389 YSOs vs. 1269 AGBs, i.e. 
52.2\,\% YSOs). We expect that the total sample of 7101 objects has a comparable 
YSO fraction of 52\,--\,53\,\%, i.e. the sample contains 3700-3800 YSOs.

The contamination among the burst candidates is significantly lower. In the
subsamples YSO1, YSO2, AGB1, AGB2, as defined above, there are 27 burst
candidates, with brightness increase of $>1.0$\,mag in the 3.6 and 4.5$\,\mu$m
bands. 17 of them are from subsample YSO1 and 10 from YSO2, none from the AGB
samples. Despite making up almost half of the entire sample, there are no AGB
stars which are highly variable. Thus, it is likely that the fraction of AGB
stars among the 77 burst candidates in the total sample is negligible. More
quantitatively, we expect that the upper limit for the fraction of AGB stars
among burst candidates is 4\,\% (1/27), which corresponds to 3 objects in
the sample of 77 burst candidates.

\newcommand\aj{AJ} 
\newcommand\actaa{AcA} 
\newcommand\araa{ARA\&A} 
\newcommand\apj{ApJ} 
\newcommand\apjl{ApJ} 
\newcommand\apjs{ApJS} 
\newcommand\aap{A\&A} 
\newcommand\aapr{A\&A~Rev.} 
\newcommand\aaps{A\&AS} 
\newcommand\mnras{MNRAS} 
\newcommand\pasa{PASA} 
\newcommand\pasp{PASP} 
\newcommand\pasj{PASJ} 
\newcommand\solphys{Sol.~Phys.} 
\newcommand\nat{Nature} 
\newcommand\bain{Bulletin of the Astronomical Institutes of the Netherlands}
\newcommand\memsai{Mem. Soc. Astron. Ital.}

\bibliographystyle{mn2e}
\bibliography{aleksbib}

\begin{thebibliography}{}

\bibitem[\protect\citeauthoryear{{Alves de Oliveira} \& {Casali}}{{Alves de
  Oliveira} \& {Casali}}{2008}]{2008A&A...485..155A}
{Alves de Oliveira} C.,  {Casali} M.,  2008, \aap, 485, 155

\bibitem[\protect\citeauthoryear{{Alves de Oliveira}, {Moraux}, {Bouvier} \&
  {Bouy}}{{Alves de Oliveira} et~al.}{2012}]{2012A&A...539A.151A}
{Alves de Oliveira} C.,  {Moraux} E.,  {Bouvier} J.,    {Bouy} H.,  2012, \aap,
  539, A151

\bibitem[\protect\citeauthoryear{{Armitage}, {Livio} \& {Pringle}}{{Armitage}
  et~al.}{2001}]{2001MNRAS.324..705A}
{Armitage} P.~J.,  {Livio} M.,    {Pringle} J.~E.,  2001, \mnras, 324, 705

\bibitem[\protect\citeauthoryear{{Bally}, {Reipurth} \& {Davis}}{{Bally}
  et~al.}{2007}]{2007prpl.conf..215B}
{Bally} J.,  {Reipurth} B.,    {Davis} C.~J.,  2007, Protostars and Planets V,
  pp 215--230

\bibitem[\protect\citeauthoryear{{Baraffe}, {Chabrier}, {Allard} \&
  {Hauschildt}}{{Baraffe} et~al.}{1998}]{1998A&A...337..403B}
{Baraffe} I.,  {Chabrier} G.,  {Allard} F.,    {Hauschildt} P.~H.,  1998, \aap,
  337, 403

\bibitem[\protect\citeauthoryear{{Barsony}, {Kenyon}, {Lada} \&
  {Teuben}}{{Barsony} et~al.}{1997}]{1997ApJS..112..109B}
{Barsony} M.,  {Kenyon} S.~J.,  {Lada} E.~A.,    {Teuben} P.~J.,  1997, \apjs,
  112, 109

\bibitem[\protect\citeauthoryear{{Bell} \& {Lin}}{{Bell} \&
  {Lin}}{1994}]{1994ApJ...427..987B}
{Bell} K.~R.,  {Lin} D.~N.~C.,  1994, \apj, 427, 987

\bibitem[\protect\citeauthoryear{{Benjamin}, {Churchwell}, {Babler}, {Bania},
  {Clemens}, {Cohen}, {Dickey}, {Indebetouw}, {Jackson}, {Kobulnicky} \& et
  al.}{{Benjamin} et~al.}{2003}]{2003PASP..115..953B}
{Benjamin} R.~A.,  {Churchwell} E.,  {Babler} B.~L.,  {Bania} T.~M.,  {Clemens}
  D.~P.,  {Cohen} M.,  {Dickey} J.~M.,  {Indebetouw} R.,  {Jackson} J.~M.,
  {Kobulnicky} H.~A.,    et al. 2003, \pasp, 115, 953

\bibitem[\protect\citeauthoryear{{Boley} \& {Durisen}}{{Boley} \&
  {Durisen}}{2008}]{2008ApJ...685.1193B}
{Boley} A.~C.,  {Durisen} R.~H.,  2008, \apj, 685, 1193

\bibitem[\protect\citeauthoryear{{Bonnell} \& {Bastien}}{{Bonnell} \&
  {Bastien}}{1992}]{1992ApJ...401L..31B}
{Bonnell} I.,  {Bastien} P.,  1992, \apjl, 401, L31

\bibitem[\protect\citeauthoryear{{Caratti o Garatti}, {Garcia Lopez}, {Scholz},
  {Giannini}, {Eisl{\"o}ffel}, {Nisini}, {Massi}, {Antoniucci} \&
  {Ray}}{{Caratti o Garatti} et~al.}{2011}]{2011A&A...526L...1C}
{Caratti o Garatti} A.,  {Garcia Lopez} R.,  {Scholz} A.,  {Giannini} T.,
  {Eisl{\"o}ffel} J.,  {Nisini} B.,  {Massi} F.,  {Antoniucci} S.,    {Ray}
  T.~P.,  2011, \aap, 526, L1+

\bibitem[\protect\citeauthoryear{{Carpenter}, {Hillenbrand} \&
  {Skrutskie}}{{Carpenter} et~al.}{2001}]{2001AJ....121.3160C}
{Carpenter} J.~M.,  {Hillenbrand} L.~A.,    {Skrutskie} M.~F.,  2001, \aj, 121,
  3160

\bibitem[\protect\citeauthoryear{{Churchwell}, {Babler}, {Meade}, {Whitney},
  {Benjamin}, {Indebetouw}, {Cyganowski}, {Robitaille}, {Povich}, {Watson} \&
  {Bracker}}{{Churchwell} et~al.}{2009}]{2009PASP..121..213C}
{Churchwell} E.,  {Babler} B.~L.,  {Meade} M.~R.,  {Whitney} B.~A.,  {Benjamin}
  R.,  {Indebetouw} R.,  {Cyganowski} C.,  {Robitaille} T.~P.,  {Povich} M.,
  {Watson} C.,    {Bracker} S.,  2009, \pasp, 121, 213

\bibitem[\protect\citeauthoryear{{Clarke}, {Lodato}, {Melnikov} \&
  {Ibrahimov}}{{Clarke} et~al.}{2005}]{2005MNRAS.361..942C}
{Clarke} C.,  {Lodato} G.,  {Melnikov} S.~Y.,    {Ibrahimov} M.~A.,  2005,
  \mnras, 361, 942

\bibitem[\protect\citeauthoryear{{Covey}, {Hillenbrand}, {Miller}, {Poznanski},
  {Cenko}, {Silverman}, {Bloom}, {Kasliwal}, {Fischer}, {Rayner}, {Rebull},
  {Butler}, {Filippenko}, {Law}, {Ofek}, {Ag{\"u}eros}, {Dekany} \& et
  al.}{{Covey} et~al.}{2011}]{2011AJ....141...40C}
{Covey} K.~R.,  {Hillenbrand} L.~A.,  {Miller} A.~A.,  {Poznanski} D.,  {Cenko}
  S.~B.,  {Silverman} J.~M.,  {Bloom} J.~S.,  {Kasliwal} M.~M.,  {Fischer} W.,
  {Rayner} J.,  {Rebull} L.~M.,  {Butler} N.~R.,  {Filippenko} A.~V.,  {Law}
  N.~M.,  {Ofek} E.~O.,  {Ag{\"u}eros} M.,  {Dekany} R.~G.,    et al. 2011,
  \aj, 141, 40

\bibitem[\protect\citeauthoryear{{Dahm} \& {Simon}}{{Dahm} \&
  {Simon}}{2005}]{2005AJ....129..829D}
{Dahm} S.~E.,  {Simon} T.,  2005, \aj, 129, 829

\bibitem[\protect\citeauthoryear{{Dunham} \& {Vorobyov}}{{Dunham} \&
  {Vorobyov}}{2012}]{2012ApJ...747...52D}
{Dunham} M.~M.,  {Vorobyov} E.~I.,  2012, \apj, 747, 52

\bibitem[\protect\citeauthoryear{{Evans}, {Dunham}, {J{\o}rgensen}, {Enoch},
  {Mer{\'{\i}}n}, {van Dishoeck}, {Alcal{\'a}}, {Myers}, {Stapelfeldt} \& {et
  al.}}{{Evans} et~al.}{2009}]{2009ApJS..181..321E}
{Evans} N.~J.,  {Dunham} M.~M.,  {J{\o}rgensen} J.~K.,  {Enoch} M.~L.,
  {Mer{\'{\i}}n} B.,  {van Dishoeck} E.~F.,  {Alcal{\'a}} J.~M.,  {Myers}
  P.~C.,  {Stapelfeldt} K.~R.,    {et al.} 2009, \apjs, 181, 321

\bibitem[\protect\citeauthoryear{{Evans} II, {Balkum}, {Levreault}, {Hartmann}
  \& {Kenyon}}{{Evans} et~al.}{1994}]{1994ApJ...424..793E}
{Evans} II N.~J.,  {Balkum} S.,  {Levreault} R.~M.,  {Hartmann} L.,    {Kenyon}
  S.,  1994, \apj, 424, 793

\bibitem[\protect\citeauthoryear{{Flaherty}, {Muzerolle}, {Rieke}, {Gutermuth},
  {Balog}, {Herbst}, {Megeath} \& {Kun}}{{Flaherty}
  et~al.}{2012}]{2012ApJ...748...71F}
{Flaherty} K.~M.,  {Muzerolle} J.,  {Rieke} G.,  {Gutermuth} R.,  {Balog} Z.,
  {Herbst} W.,  {Megeath} S.~T.,    {Kun} M.,  2012, \apj, 748, 71

\bibitem[\protect\citeauthoryear{{Forgan} \& {Rice}}{{Forgan} \&
  {Rice}}{2010}]{2010MNRAS.402.1349F}
{Forgan} D.,  {Rice} K.,  2010, \mnras, 402, 1349

\bibitem[\protect\citeauthoryear{{Froebrich}, {Davis}, {Ioannidis}, {Gledhill},
  {Takami}, {Chrysostomou}, {Drew}, {Eisl{\"o}ffel}, {Gosling}, {Gredel},
  {Hatchell}, {Hodapp}, {Kumar}, {Lucas}, {Matthews} \& et al.}{{Froebrich}
  et~al.}{2011}]{2011MNRAS.413..480F}
{Froebrich} D.,  {Davis} C.~J.,  {Ioannidis} G.,  {Gledhill} T.~M.,  {Takami}
  M.,  {Chrysostomou} A.,  {Drew} J.,  {Eisl{\"o}ffel} J.,  {Gosling} A.,
  {Gredel} R.,  {Hatchell} J.,  {Hodapp} K.~W.,  {Kumar} M.~S.~N.,  {Lucas}
  P.~W.,  {Matthews} H.,    et al. 2011, \mnras, 413, 480

\bibitem[\protect\citeauthoryear{{Gutermuth}, {Megeath}, {Myers}, {Allen},
  {Pipher} \& {Fazio}}{{Gutermuth} et~al.}{2009}]{2009ApJS..184...18G}
{Gutermuth} R.~A.,  {Megeath} S.~T.,  {Myers} P.~C.,  {Allen} L.~E.,  {Pipher}
  J.~L.,    {Fazio} G.~G.,  2009, \apjs, 184, 18

\bibitem[\protect\citeauthoryear{{Hartmann} \& {Kenyon}}{{Hartmann} \&
  {Kenyon}}{1996}]{1996ARA&A..34..207H}
{Hartmann} L.,  {Kenyon} S.~J.,  1996, \araa, 34, 207

\bibitem[\protect\citeauthoryear{{Herbig}}{{Herbig}}{1977}]{1977ApJ...217..693%
H}
{Herbig} G.~H.,  1977, \apj, 217, 693

\bibitem[\protect\citeauthoryear{{Herbig}}{{Herbig}}{2007}]{2007AJ....133.2679%
H}
{Herbig} G.~H.,  2007, \aj, 133, 2679

\bibitem[\protect\citeauthoryear{{Ioannidis} \& {Froebrich}}{{Ioannidis} \&
  {Froebrich}}{2012}]{2012arXiv1206.5095I}
{Ioannidis} G.,  {Froebrich} D.,  2012, ArXiv e-prints

\bibitem[\protect\citeauthoryear{{K{\'o}sp{\'a}l}, {{\'A}brah{\'a}m},
  {Acosta-Pulido}, {Ar{\'e}valo Morales}, {Carnerero}, {Elek}, {Kelemen},
  {Kun}, {P{\'a}l}, {Szak{\'a}ts} \& {Vida}}{{K{\'o}sp{\'a}l}
  et~al.}{2011}]{2011A&A...527A.133K}
{K{\'o}sp{\'a}l} {\'A}.,  {{\'A}brah{\'a}m} P.,  {Acosta-Pulido} J.~A.,
  {Ar{\'e}valo Morales} M.~J.,  {Carnerero} M.~I.,  {Elek} E.,  {Kelemen} J.,
  {Kun} M.,  {P{\'a}l} A.,  {Szak{\'a}ts} R.,    {Vida} K.,  2011, \aap, 527,
  A133+

\bibitem[\protect\citeauthoryear{{Lada} \& {Lada}}{{Lada} \&
  {Lada}}{2003}]{2003ARA&A..41...57L}
{Lada} C.~J.,  {Lada} E.~A.,  2003, \araa, 41, 57

\bibitem[\protect\citeauthoryear{{Lodato} \& {Clarke}}{{Lodato} \&
  {Clarke}}{2004}]{2004MNRAS.353..841L}
{Lodato} G.,  {Clarke} C.~J.,  2004, \mnras, 353, 841

\bibitem[\protect\citeauthoryear{{Martin} \& {Lubow}}{{Martin} \&
  {Lubow}}{2011}]{2011ApJ...740L...6M}
{Martin} R.~G.,  {Lubow} S.~H.,  2011, \apjl, 740, L6

\bibitem[\protect\citeauthoryear{{Miller}, {Hillenbrand}, {Covey}, {Poznanski},
  {Silverman}, {Kleiser}, {Rojas-Ayala}, {Muirhead}, {Cenko}, {Bloom},
  {Kasliwal}, {Filippenko}, {Law}, {Ofek}, {Dekany}, {Rahmer} \& {et
  al.}}{{Miller} et~al.}{2011}]{2011ApJ...730...80M}
{Miller} A.~A.,  {Hillenbrand} L.~A.,  {Covey} K.~R.,  {Poznanski} D.,
  {Silverman} J.~M.,  {Kleiser} I.~K.~W.,  {Rojas-Ayala} B.,  {Muirhead} P.~S.,
   {Cenko} S.~B.,  {Bloom} J.~S.,  {Kasliwal} M.~M.,  {Filippenko} A.~V.,
  {Law} N.~M.,  {Ofek} E.~O.,  {Dekany} R.~G.,  {Rahmer} G.,    {et al.} 2011,
  \apj, 730, 80

\bibitem[\protect\citeauthoryear{{Morales-Calder{\'o}n}, {Stauffer},
  {Hillenbrand}, {Gutermuth}, {Song}, {Rebull}, {Plavchan}, {Carpenter},
  {Whitney}, {Covey}, {Alves de Oliveira}, {Winston}, {McCaughrean} \& {et
  al.}}{{Morales-Calder{\'o}n} et~al.}{2011}]{2011ApJ...733...50M}
{Morales-Calder{\'o}n} M.,  {Stauffer} J.~R.,  {Hillenbrand} L.~A.,
  {Gutermuth} R.,  {Song} I.,  {Rebull} L.~M.,  {Plavchan} P.,  {Carpenter}
  J.~M.,  {Whitney} B.~A.,  {Covey} K.,  {Alves de Oliveira} C.,  {Winston} E.,
   {McCaughrean} M.~J.,    {et al.} 2011, \apj, 733, 50

\bibitem[\protect\citeauthoryear{{Natta}, {Testi} \& {Randich}}{{Natta}
  et~al.}{2006}]{2006A&A...452..245N}
{Natta} A.,  {Testi} L.,    {Randich} S.,  2006, \aap, 452, 245

\bibitem[\protect\citeauthoryear{{Offner} \& {McKee}}{{Offner} \&
  {McKee}}{2011}]{2011ApJ...736...53O}
{Offner} S.~S.~R.,  {McKee} C.~F.,  2011, \apj, 736, 53

\bibitem[\protect\citeauthoryear{{Pfalzner}}{{Pfalzner}}{2008}]{2008A&A...492.%
.735P}
{Pfalzner} S.,  2008, \aap, 492, 735

\bibitem[\protect\citeauthoryear{{Rebull}, {Guieu}, {Stauffer}, {Hillenbrand},
  {Noriega-Crespo}, {Stapelfeldt}, {Carey}, {Carpenter}, {Cole}, {Padgett},
  {Strom} \& {Wolff}}{{Rebull} et~al.}{2011}]{2011ApJS..193...25R}
{Rebull} L.~M.,  {Guieu} S.,  {Stauffer} J.~R.,  {Hillenbrand} L.~A.,
  {Noriega-Crespo} A.,  {Stapelfeldt} K.~R.,  {Carey} S.~J.,  {Carpenter}
  J.~M.,  {Cole} D.~M.,  {Padgett} D.~L.,  {Strom} S.~E.,    {Wolff} S.~C.,
  2011, \apjs, 193, 25

\bibitem[\protect\citeauthoryear{{Rebull}, {Padgett}, {McCabe}, {Hillenbrand},
  {Stapelfeldt}, {Noriega-Crespo}, {Carey}, {Brooke}, {Huard}, {Terebey},
  {Audard}, {Monin}, {Fukagawa}, {G{\"u}del} \& et al.}{{Rebull}
  et~al.}{2010}]{2010ApJS..186..259R}
{Rebull} L.~M.,  {Padgett} D.~L.,  {McCabe} C.-E.,  {Hillenbrand} L.~A.,
  {Stapelfeldt} K.~R.,  {Noriega-Crespo} A.,  {Carey} S.~J.,  {Brooke} T.,
  {Huard} T.,  {Terebey} S.,  {Audard} M.,  {Monin} J.-L.,  {Fukagawa} M.,
  {G{\"u}del} M.,    et al. 2010, \apjs, 186, 259

\bibitem[\protect\citeauthoryear{{Reipurth}}{{Reipurth}}{1985}]{1985A&A...143.%
.435R}
{Reipurth} B.,  1985, \aap, 143, 435

\bibitem[\protect\citeauthoryear{{Reipurth} \& {Aspin}}{{Reipurth} \&
  {Aspin}}{2010}]{2010vaoa.conf...19R}
{Reipurth} B.,  {Aspin} C.,  2010, in {Harutyunian} H.~A.,  {Mickaelian} A.~M.,
    {Terzian} Y.,  eds, Evolution of Cosmic Objects through their Physical
  Activity {FUors and Early Stellar Evolution}.
pp 19--38

\bibitem[\protect\citeauthoryear{{Reipurth}, {Aspin} \& {Herbig}}{{Reipurth}
  et~al.}{2012}]{2012ApJ...748L...5R}
{Reipurth} B.,  {Aspin} C.,    {Herbig} G.~H.,  2012, \apjl, 748, L5

\bibitem[\protect\citeauthoryear{{Robitaille}, {Meade}, {Babler}, {Whitney},
  {Johnston}, {Indebetouw}, {Cohen}, {Povich}, {Sewilo}, {Benjamin} \&
  {Churchwell}}{{Robitaille} et~al.}{2008}]{2008AJ....136.2413R}
{Robitaille} T.~P.,  {Meade} M.~R.,  {Babler} B.~L.,  {Whitney} B.~A.,
  {Johnston} K.~G.,  {Indebetouw} R.,  {Cohen} M.,  {Povich} M.~S.,  {Sewilo}
  M.,  {Benjamin} R.~A.,    {Churchwell} E.,  2008, \aj, 136, 2413

\bibitem[\protect\citeauthoryear{{Robitaille}, {Whitney}, {Indebetouw}, {Wood}
  \& {Denzmore}}{{Robitaille} et~al.}{2006}]{2006ApJS..167..256R}
{Robitaille} T.~P.,  {Whitney} B.~A.,  {Indebetouw} R.,  {Wood} K.,
  {Denzmore} P.,  2006, \apjs, 167, 256

\bibitem[\protect\citeauthoryear{{Saito}, {Hempel}, {Minniti}, {Lucas},
  {Rejkuba}, {Toledo}, {Gonzalez}, {Alonso-Garc{\'{\i}}a}, {Irwin},
  {Gonzalez-Solares}, {Hodgkin}, {Lewis}, {Cross}, {Ivanov}, {Kerins} \& et
  al.}{{Saito} et~al.}{2012}]{2012A&A...537A.107S}
{Saito} R.~K.,  {Hempel} M.,  {Minniti} D.,  {Lucas} P.~W.,  {Rejkuba} M.,
  {Toledo} I.,  {Gonzalez} O.~A.,  {Alonso-Garc{\'{\i}}a} J.,  {Irwin} M.~J.,
  {Gonzalez-Solares} E.,  {Hodgkin} S.~T.,  {Lewis} J.~R.,  {Cross} N.,
  {Ivanov} V.~D.,  {Kerins} E.,    et al. 2012, \aap, 537, A107

\bibitem[\protect\citeauthoryear{{Sandell} \& {Weintraub}}{{Sandell} \&
  {Weintraub}}{2001}]{2001ApJS..134..115S}
{Sandell} G.,  {Weintraub} D.~A.,  2001, \apjs, 134, 115

\bibitem[\protect\citeauthoryear{{Scholz}}{{Scholz}}{2012}]{2012MNRAS.420.1495%
S}
{Scholz} A.,  2012, \mnras, 420, 1495

\bibitem[\protect\citeauthoryear{{Scholz}, {Jayawardhana} \& {Wood}}{{Scholz}
  et~al.}{2006}]{2006ApJ...645.1498S}
{Scholz} A.,  {Jayawardhana} R.,    {Wood} K.,  2006, \apj, 645, 1498

\bibitem[\protect\citeauthoryear{{Stamatellos}, {Whitworth} \&
  {Hubber}}{{Stamatellos} et~al.}{2011}]{2011ApJ...730...32S}
{Stamatellos} D.,  {Whitworth} A.~P.,    {Hubber} D.~A.,  2011, \apj, 730, 32

\bibitem[\protect\citeauthoryear{{Sung}, {Stauffer} \& {Bessell}}{{Sung}
  et~al.}{2009}]{2009AJ....138.1116S}
{Sung} H.,  {Stauffer} J.~R.,    {Bessell} M.~S.,  2009, \aj, 138, 1116

\bibitem[\protect\citeauthoryear{{Vorobyov} \& {Basu}}{{Vorobyov} \&
  {Basu}}{2005}]{2005ApJ...633L.137V}
{Vorobyov} E.~I.,  {Basu} S.,  2005, \apjl, 633, L137

\bibitem[\protect\citeauthoryear{{Welin}}{{Welin}}{1971}]{1971A&A....12..312W}
{Welin} G.,  1971, \aap, 12, 312

\bibitem[\protect\citeauthoryear{{Wright}, {Eisenhardt}, {Mainzer}, {Ressler},
  {Cutri}, {Jarrett}, {Kirkpatrick}, {Padgett}, {McMillan}, {Skrutskie},
  {Stanford}, {Cohen}, {Walker}, {Mather} \& {et al.}}{{Wright}
  et~al.}{2010}]{2010AJ....140.1868W}
{Wright} E.~L.,  {Eisenhardt} P.~R.~M.,  {Mainzer} A.~K.,  {Ressler} M.~E.,
  {Cutri} R.~M.,  {Jarrett} T.,  {Kirkpatrick} J.~D.,  {Padgett} D.,
  {McMillan} R.~S.,  {Skrutskie} M.,  {Stanford} S.~A.,  {Cohen} M.,  {Walker}
  R.~G.,  {Mather} J.~C.,    {et al.} 2010, \aj, 140, 1868

\bibitem[\protect\citeauthoryear{{Zhu}, {Hartmann}, {Gammie} \&
  {McKinney}}{{Zhu} et~al.}{2009}]{2009ApJ...701..620Z}
{Zhu} Z.,  {Hartmann} L.,  {Gammie} C.,    {McKinney} J.~C.,  2009, \apj, 701,
  620

\end{thebibliography}

\label{lastpage}

\end{document}